\documentclass[10pt, pra, aps, twocolumn, superscriptaddress, citeautoscript]{revtex4-1}
\usepackage{color}
\usepackage{graphicx}
\usepackage{hyperref}
\usepackage{amssymb, amstext, amsmath}

\DeclareGraphicsExtensions{{.eps},{.png},{.pdf}}

\def\kB{k_{\text{B}}}

\begin{document}

\title{Observation of two-dimensional Anderson localisation of ultracold atoms}

\author{Donald H. White}
    \altaffiliation[Present address:\ ]{Department of Applied Physics, Waseda University, Shinjuku, Tokyo, Japan}
    \affiliation{Dodd-Walls Centre for Photonic and Quantum Technologies, New Zealand}
    \affiliation{Department of Physics, University of Auckland, Auckland, New Zealand}
\author{Thomas A.~Haase}
    \thanks{D.\,H.\,W. and T.\,A.\,H. contributed equally to this work.}
    \affiliation{Dodd-Walls Centre for Photonic and Quantum Technologies, New Zealand}
    \affiliation{Department of Physics, University of Auckland, Auckland, New Zealand}
\author{Dylan J.~Brown}
    \affiliation{Dodd-Walls Centre for Photonic and Quantum Technologies, New Zealand}
    \affiliation{Department of Physics, University of Auckland, Auckland, New Zealand}
\author{Maarten D.~Hoogerland}
    \affiliation{Dodd-Walls Centre for Photonic and Quantum Technologies, New Zealand}
    \affiliation{Department of Physics, University of Auckland, Auckland, New Zealand}

\author{\\ Mojdeh S.~Najafabadi}
    \affiliation{Dodd-Walls Centre for Photonic and Quantum Technologies, New Zealand}
    \affiliation{Department of Physics, University of Otago, Dunedin, New Zealand}
\author{John L.~Helm}
    \affiliation{Dodd-Walls Centre for Photonic and Quantum Technologies, New Zealand}
    \affiliation{Department of Physics, University of Otago, Dunedin, New Zealand}
\author{Christopher Gies}
    \affiliation{Institut f{\"u}r Theoretische Physik, Universit{\"a}t Bremen, Bremen, Germany}
\author{Daniel Schumayer}
    \affiliation{Dodd-Walls Centre for Photonic and Quantum Technologies, New Zealand}
    \affiliation{Department of Physics, University of Otago, Dunedin, New Zealand}
\author{David A.~W.~Hutchinson}
    \affiliation{Dodd-Walls Centre for Photonic and Quantum Technologies, New Zealand}
    \affiliation{Department of Physics, University of Otago, Dunedin, New Zealand}

\maketitle

\section{Abstract}

{\textbf{
Anderson localisation ---the inhibition of wave propagation in disordered
media--- is a surprising interference phenomenon which is particularly intriguing
in two-dimensional (2D) systems. While an ideal, non-interacting 2D system of
infinite size is always localised, the localisation length-scale may be too
large to be unambiguously observed in an experiment. In this sense, 2D is a
marginal dimension between one-dimension, where all states are strongly
localised, and three-dimensions, where a well-defined phase transition between localisation and delocalisation
exists as the energy is increased. Here we report the results of an experiment measuring the 2D transport of ultracold atoms between two reservoirs, which are connected by a channel containing pointlike disorder. The design overcomes many of the technical challenges that have hampered observation of localisation in previous works. We experimentally observe exponential localisation in a 2D ultracold atom system.
}}

\section{Introduction}

Anderson localisation \cite{Anderson1958} is a phenomenon resulting from wave
interference between multiple propagation paths, and has been observed in a
variety of wave systems \cite{Mott1969, Lee1985, Weaver1990, McCall1991,
Wiersma1997, Genack1997, Weiland1999, Storzer2006, Topolancik2007, Schwartz2007,
Hu2008, Chabe2008, Riboli2011, Sperling2012, Lopez2012, Manai2015, Ying2016}.
While it is a single-particle phenomenon, its nature is affected by numerous factors including interparticle interactions
\cite{Fishman2012, Shepelyansky1993}, dimensionality \cite{Abrahams1979},
time-reversal symmetry \cite{Bergmann1984}, spin-orbit coupling
\cite{Bergmann1982}, and the microscopic nature of the disorder
\cite{Piraud2012}. A full understanding of the physics of Anderson localisation demands experimental control of these parameters. Ultracold atoms have proven to be among the
cleanest and most controllable of all quantum mechanical systems
\cite{Georgescu2014}, and have thus provided a natural avenue for modern
experiments on Anderson localisation.

Careful experiments in 1D with weakly-~\cite{Aspect2008} and non-interacting~\cite{Inguscio2008} atoms expanding in a waveguide
containing optically-generated disorder allowed for unambiguous observation of
Anderson localisation. These were followed
by experiments demonstrating Anderson localisation in 3D \cite{deMarco2011,
Aspect2012}, and by studies of the metal-insulator transition
\cite{Semeghini2015}.

In parallel to this, experiments with cold atoms in 2D have shown behaviours
characteristic of weak localisation \cite{Vincent2010, Jendrzejewski2012,
Mueller2015}. However, unambiguous observation of Anderson localisation in 2D
real-space cold atom systems has, to our knowledge, not previously been
observed. This has been due to two main challenges. First, the localisation
length in 2D depends exponentially on the particle energy \cite{Lee1985,Muller2005}: for
experimentally feasible particle energies, observing localisation requires large
systems ($>100\mu\text{m} \times 100\mu$m) even for ultracold atoms. The
optically disordered potential landscapes must have high optical resolution over
the entire domain, because the scatterer size must be smaller than the atomic de
Broglie wavelength (equivalently, the spatial Fourier components of disorder
must exceed the majority of atomic momenta). Secondly, while optical speckle
patterns provide appropriate disorder for 1D and 3D systems, the statistics of
optical speckle are problematic in 2D due to the high classical percolation
threshold
\cite{Morong2015}. Observing Anderson localisation in 2D on reasonable
length-scales, therefore, requires relatively strong scattering, and this leads
to difficulty in distinguishing localisation effects from classical trapping;
low energy particles have the shortest localisation lengths, yet they are also
trapped classically by the optical speckle. To this end, Morong and DeMarco
suggested the use of randomly positioned point scatterers \cite{Morong2015},
which allows for a tuneable percolation threshold based on the amount of
disorder, and thus allows for quantum interference effects to be effectively
isolated from trapping effects.

In this work we implement point scatterers in a 2D plane by projecting a
blue-detuned 532 nm optical pattern shaped by a spatial light modulator (SLM)
onto a flat, large-area two-dimensional trap formed from 1064\,nm light
\cite{Haase2017}. The SLM enables any arbitrary potential to be projected onto
this plane. We take advantage of this flexibility and project the outline of an
additional dumbbell-shaped container consisting of two reservoirs separated by a
channel~\cite{Eckel2016, Li2016}, with point scatterer disorder located in the
channel. Atoms from a $^{87}$Rb Bose--Einstein Condensate (BEC) are loaded into
the source reservoir, and propagate through the channel into the drain
reservoir. The transmissive nature of this experiment has four main advantages
compared to traditional expansion experiments with ultracold atoms
\cite{Sanchez-Palencia2008}. Firstly, the fraction of atoms collected in the
source and drain reservoirs provides a measurement of the effective resistance
of the disordered channel. The measurement of the atom number in a finite reservoir
provides a larger signal-to-noise ratio than is accessible with an expansion
experiment. Secondly, measuring the atom distribution within the channel enables
us to identify the onset of strong localisation as the channel density profile
becomes exponential. The two complementary measurements, of the resistance and the channel profile, provide rich information on the transport properties of the
disordered channel. Thirdly, the transmissive nature of the experiment allows us
to arbitrarily change the length and width of the atom container, and thus to
observe the atom transport on length scales both shorter and longer than the
localisation length $\xi$. Finally, in a transport experiment the Bose gas is
not in thermal equilibrium, which suppresses the formation of a Lifshits glass
\cite{Lifshits1988, Lugan2007} (the mixture of low-energy single-particle
localised states could mistakenly be identified as Anderson localisation). With
these advantages, we tune between the weak- and strong-localised regimes
\cite{Hsu1995}, and observe compelling evidence for Anderson localisation of ultracold atoms in 2D.

\begin{figure}[b!]
    \includegraphics[width=86mm]{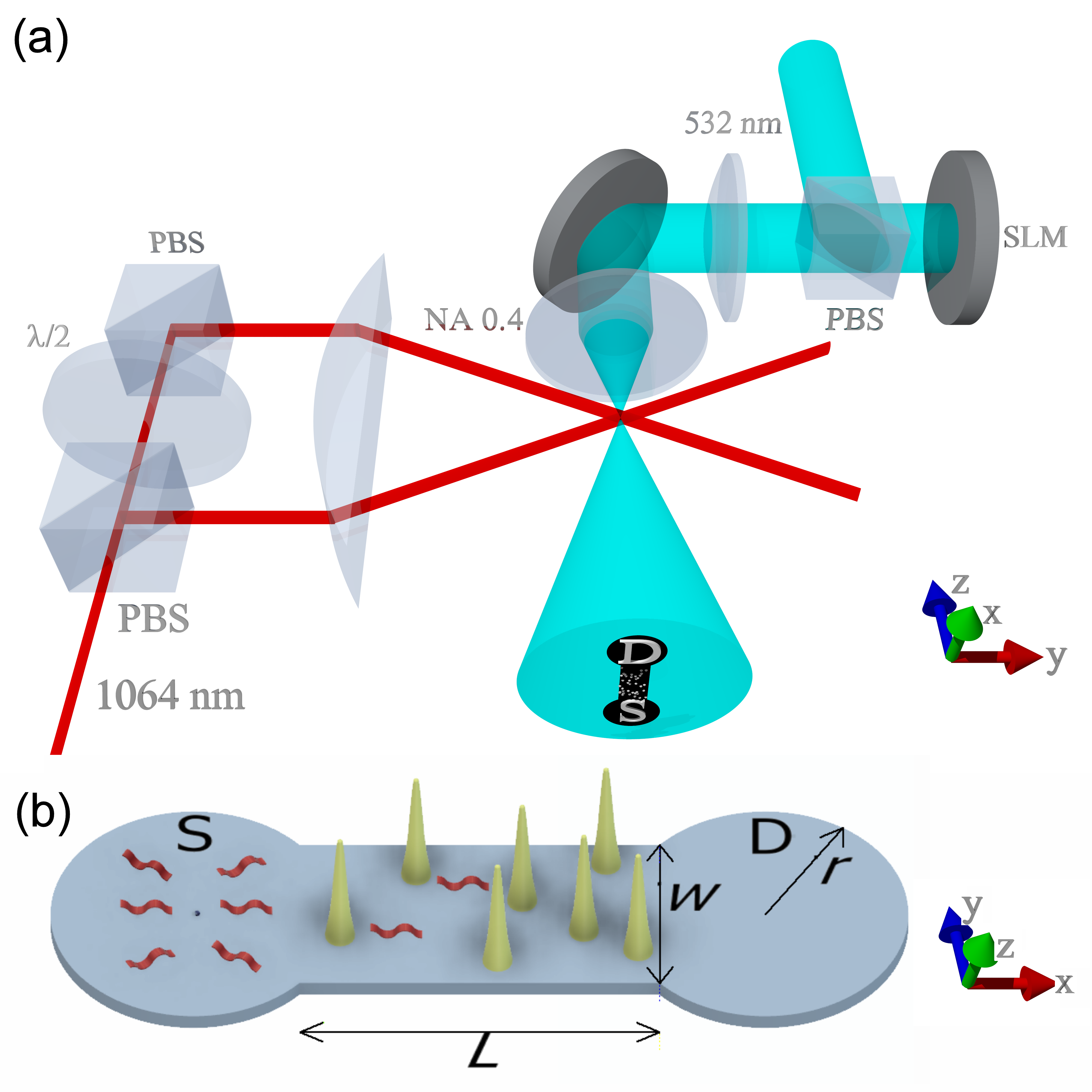}
    \caption{\label{Fig:ExperimentalSetup}
             \textbf{Experimental setup.}
             (a) The two-dimensional trap is produced by interfering two 1064~nm beams, focused with a 250\,mm focal length cylindrical lens. The beams intersect at a relative angle of $6^{\circ}$ producing horizontal pancake-like interference fringes in the vacuum chamber where the BEC is prepared in a crossed-beam CO$_2$ laser optical dipole trap. Atoms from the BEC are loaded into a single light sheet. Simultaneously, a wide and uniform
             beam of blue-detuned 532\,nm light (top right) is reflected from the spatial light modulator (SLM), with a dumbbell-shaped mask applied. Disorder is located within the channel connecting the two reservoirs of the dumbbell. The polarising beamsplitter (PBS) converts the spatial polarisation modulation of the SLM to intensity modulation, which is imaged onto the atomic plane using two lenses. The in-vacuum aspheric lens of numerical aperture 0.42 provides a resolution of 0.9~$\mu$m. An example of the dumbbell-shaped combined red and blue optical potentials at
             the atomic plane are shown in the expanded bottom right bubble. Atoms are loaded into the source (S) reservoir and propagate through the channel into the drain reservoir (D). (b) Atoms are released from the CO$_2$ laser trap at the centre of the source reservoir, and propagate as matter waves (red) into the disordered channel and drain reservoir. The radius $r$, channel length $L$ and channel width $w$ are illustrated. Point disorder within the channel is illustrated as a series of potential hills.
            }
\end{figure}

\section{Methods}

A BEC of \textsuperscript{87}Rb atoms is initially
prepared in a crossed-beam CO$_2$ laser optical dipole trap and $\sim 1.6\times
10^4$ atoms in the $\lvert F=1, m_{\mathrm{F}}=-1 \rangle$ state are loaded into a
large-area quasi-2D trap, as illustrated in Fig.~\ref{Fig:ExperimentalSetup}. The trap is created by interfering two elliptical
beams ($1.8$\,mm-to-$8$\,mm height-to-width ratio), each of 5.0\,W of 1064\,nm
light at an angle of $6^{\circ}$. The resulting light sheets are vertically
spaced by 8\,$\mu$m, while the initial diameter of the three-dimensional BEC is
$\sim 2\,\mu$m. This allows the $\sim 5$\,nK atoms to load into a single light
sheet, with characteristic trap frequencies of $\nu_{\mathrm{x}}=\nu_{\mathrm{y}}=1$\,Hz,
$\nu_{\mathrm{z}}=800$\,Hz. This geometry produces a nearly flat potential in the
horizontal dimensions, allowing near-ballistic transport with the exception of a weak long-period in-trap interference fringe~\cite{SupplementaryMaterial}.

\begin{figure*}[bth]
    \includegraphics[width=160mm]{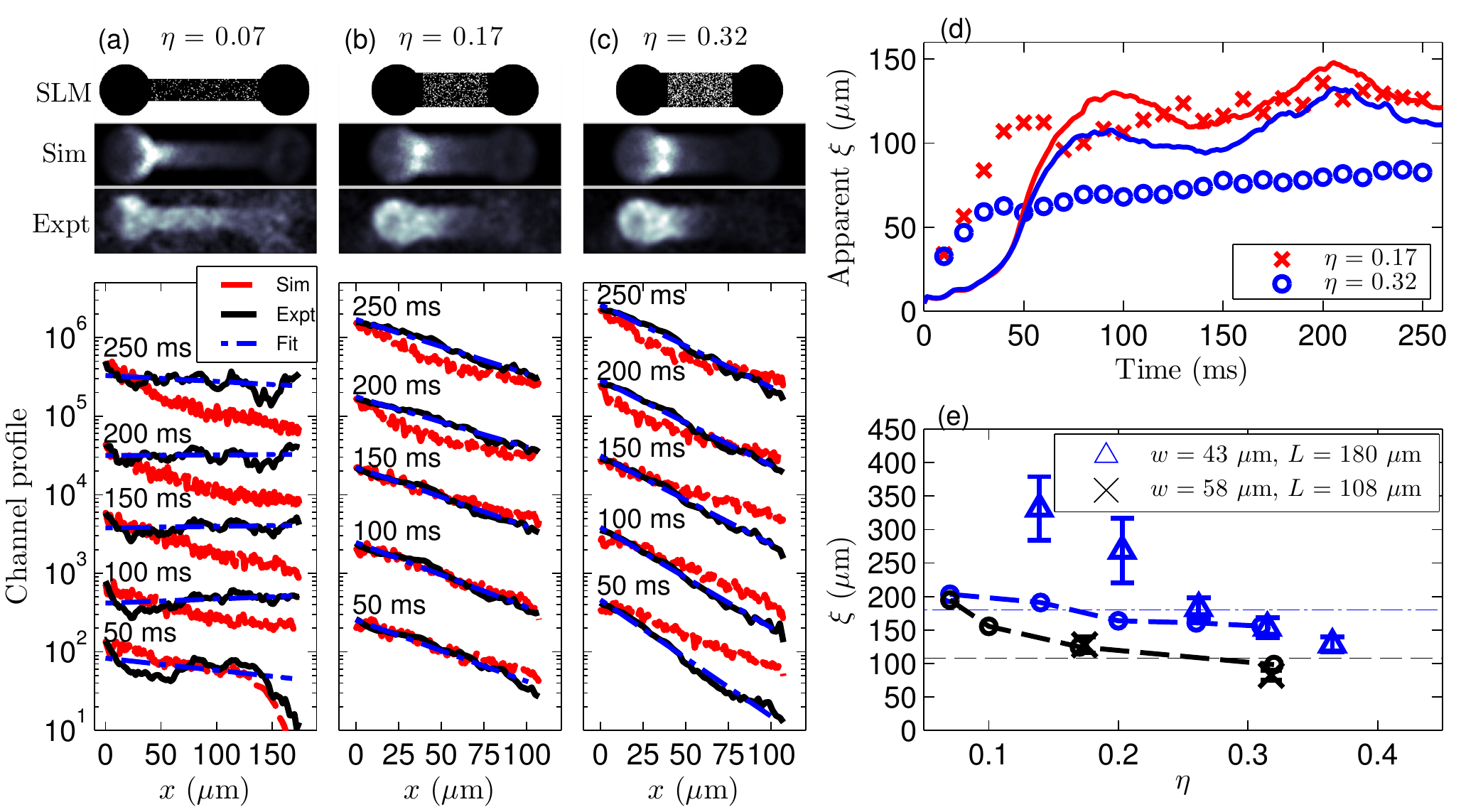}
    \caption{\label{fig:ComplexFig}
             \textbf{Observation of exponential channel density profiles.}
             (a-c) The top images in each column show the mask applied to the spatial light modulator (white indicates bright pixels). The second row of images shows the density obtained from Gross--Pitaevskii simulations after 250 ms. The third row of images show an average of three experimental absorption images after 250 ms of evolution, each with different disorder realisations. The channel density profiles
                   show semi-logarithmic snapshots of the channel density (in
                   units of atoms per 2.1~$\mu$m camera pixel length), at times (50, 100,
                   150, 200, 250) ms of time evolution, with the density integrated across the $y$-direction. Each increasing-time
                   snapshot is offset for clarity by a factor of 10. Profiles are
                   overlaid with an exponential fit to the data in blue, and with the density profiles from the GPE simulation in red. Details of the geometry are: (a) $\eta = 0.07$, $(r, L, w)=(43, 180, 36)\,
                   \mu$m; (b) $\eta = 0.17$, $(r, L,w) = (43, 108, 58)\,\mu$m;
                   (c) $\eta = 0.32$, $(r, L, w) = (43, 108, 58)\,\mu$m.
(d) The apparent localisation length is found at each time
                 evolution from the weighted exponential fit to the channel
                 profile for two values of $\eta$, with $(r, L, w) = (43, 108,
                 58)\,\mu$m. Results from GPE simulations are shown as solid
                 lines.
             (e) The localisation length is found as an average of apparent
                 localisation lengths for times 210--250 ms, for two channel
                 geometries. Numerical simulation data is also plotted (joined
                 circles). Errorbars show standard deviations obtained from three trials with different disorder realisations. Dotted lines indicate the channel lengths of $180\,\mu$m and $108\,\mu$m of the two different geometries. Note that the experimental data for $w=58~\mu$m is shown for $\eta=0.17$ and $\eta=0.32$ only.
           }
\end{figure*}
A custom optical potential, produced with an image of a $1280 \times 768$ pixel
Holoeye LC-R 720 spatial light modulator (SLM) is then projected onto the
working plane. The image is generated with blue-detuned 532\,nm light, and
imaged with an in-vacuum aspheric lens of numerical aperture 0.42 to give a
resolution of $0.9\,\mu$m. A single SLM pixel has dimensions $20\,\mu\text{m}
\times 20\,\mu$m, which with a magnification of 0.036 translates to a dimension
of $0.72\,\mu\text{m} \times 0.72\,\mu$m in
the image plane. The setup allows any custom potential to be drawn, and we image
a dumbbell-shaped boundary of two reservoirs of radius $r$ separated in the $x$
direction and linked by a channel of length $L$ and width $w$. The channel
contains customisable, point-like, optical disorder, produced by images of
randomly located blocks of $2\times 2$ SLM pixels. In the image plane these manifest as approximately Gaussian potential hills of full-width-at-half-maximum $\sigma =
1.4\,\mu$m and 5\,nK amplitude.

\begin{figure}[tb]
    \includegraphics[width=86mm]{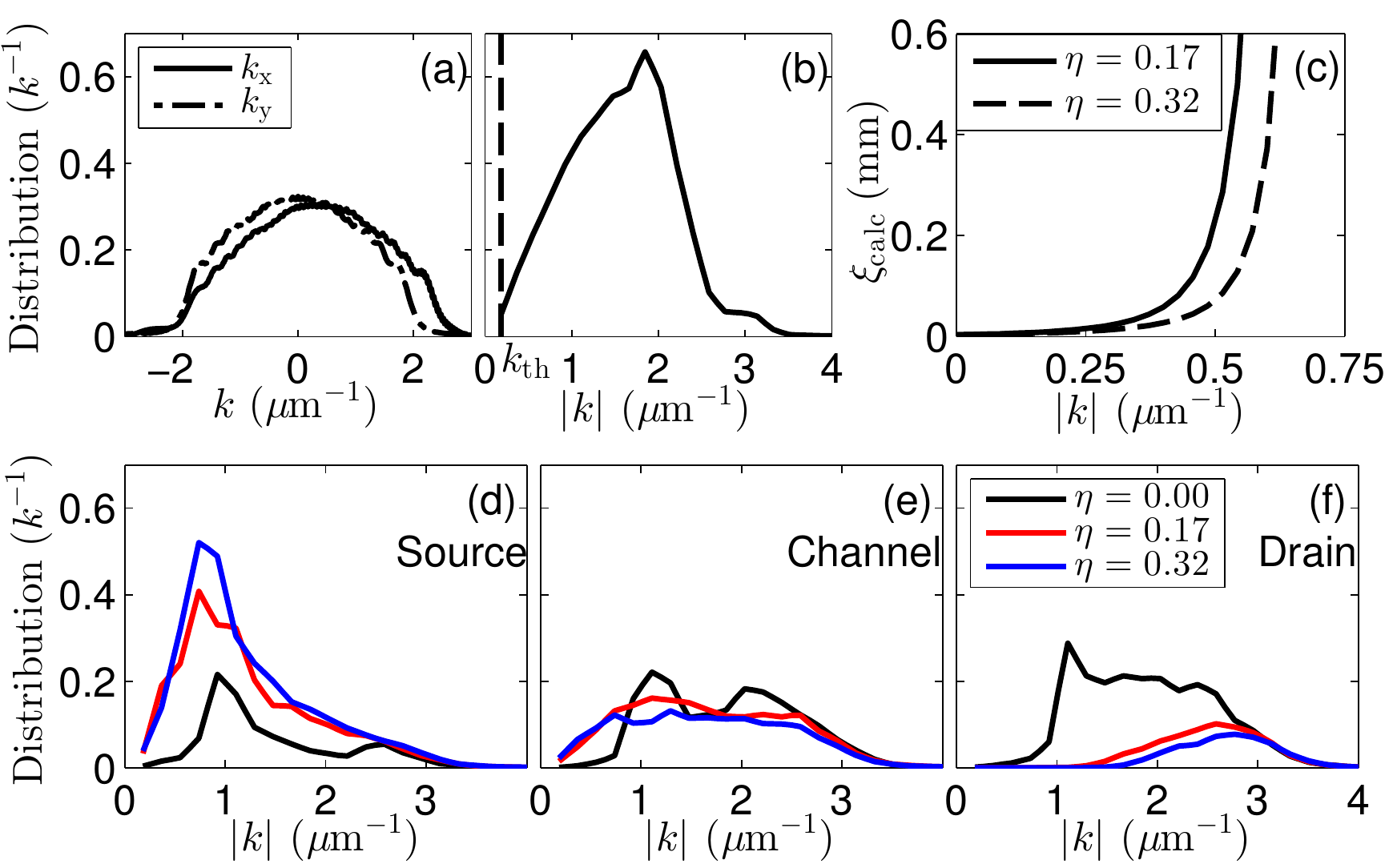}
    \caption{\label{Fig:MomentumDist}
             \textbf{Momentum distribution of atoms and the link to localisation length.}
(a) Numerical simulation showing the initial momentum distribution in the $x$ and $y$ directions, following the interaction-driven expansion after release from the harmonic trap, and prior to entering the channel (40 ms of expansion). (b) Distribution of initial absolute momentum value, with a mean of approximately 1.6~$\mu$m$^{-1}$. The classical trapping threshold is indicated as the dashed line annotated by $k_{\mathrm{th}}$. (c) Theoretical localisation length as a function of $|k|$, as calculated from Kuhn \textit{et al.} for two fill-factors ~\cite{Muller2005}. Following 250 ms of expansion in the numerical simulation, the $k$-distribution in the three dumbbell regions of a $L=108~\mu$m dumbbell is plotted for three fill-factors for (d) the source reservoir; (e) the channel; and (f) the drain reservoir. The scale is chosen such that the integral of each distribution reflects the sum of atoms in that region of the dumbbell.
            }
\end{figure}

Atoms are loaded at the centre of the source reservoir and propagate through the
channel into the drain reservoir for an expansion time $t$ after the
CO$_2$ laser crossed-beam trap is released. The atoms initially expand due to repulsive atom-atom interactions, acquiring kinetic energy and a mean wavenumber of $k\approx 1.6\,\mu$m$^{-1}$. The disorder correlation length is approximately one quarter of the de Broglie wavelength, giving the wave scattering properties which allow for Anderson
localisation, especially for atoms with energies lower than the mean energy.

Once the atoms have been loaded into the 2D trap, they are left to expand
through the channel into the second reservoir. We impart a weak slope to the
trap, giving a gravitational acceleration of $\sim 0.002$\,m/s$^2$ along the
longitudinal direction and thereby atoms acquire $\sim 0.6 \kB T$ of kinetic
energy while crossing a $150\,\mu$m channel. This linear potential
assists the transport by compensating for a weak fringing barrier~\cite{SupplementaryMaterial} at the opening
of the source reservoir and it is analogous to a weak voltage applied to an
electronic thin film in order to obtain a resistance measurement. For sufficiently weak bias, Anderson localisation is expected to be maintained~\cite{Bellaistre2017}. Data
acquisition is performed by capturing a series of absorption images, with imaging resolution of 8~$\mu$m, at different expansion
times within the dumbbell in steps of 10\,ms up to 250\,ms. Example
absorption images are shown in the `Expt' panels of Fig.~\ref{fig:ComplexFig}(a)-(c).
For each fill-factor the experiment is repeated three times, each time with a
different disorder realisation to perform configurational averaging.

The disorder is characterised by its fill-factor $\eta$, defined as: $\eta = A_{\text{disorder}}/A_{\text{channel}} = n \sigma^2$, where $n$
is the density of scatterers and $\sigma=1.4\,\mu$m is the effective scatterer
width. Equivalently, $\eta$ is the fraction of bright pixels within the channel
displayed by the SLM. Note that the classical percolation threshold of point
scatterer disorder is negligible for $\eta\lesssim 0.06$ and remains below that
of the optical speckle up to $\eta\lesssim 0.35$ \cite{Morong2015}. In referring to Ref.~\cite{Morong2015}, note that our definition of fill-factor differs by a factor of 2, i.e. $\eta = 2nw^2$.

\begin{figure}[ht]
    \centering
    \includegraphics[width=250px]{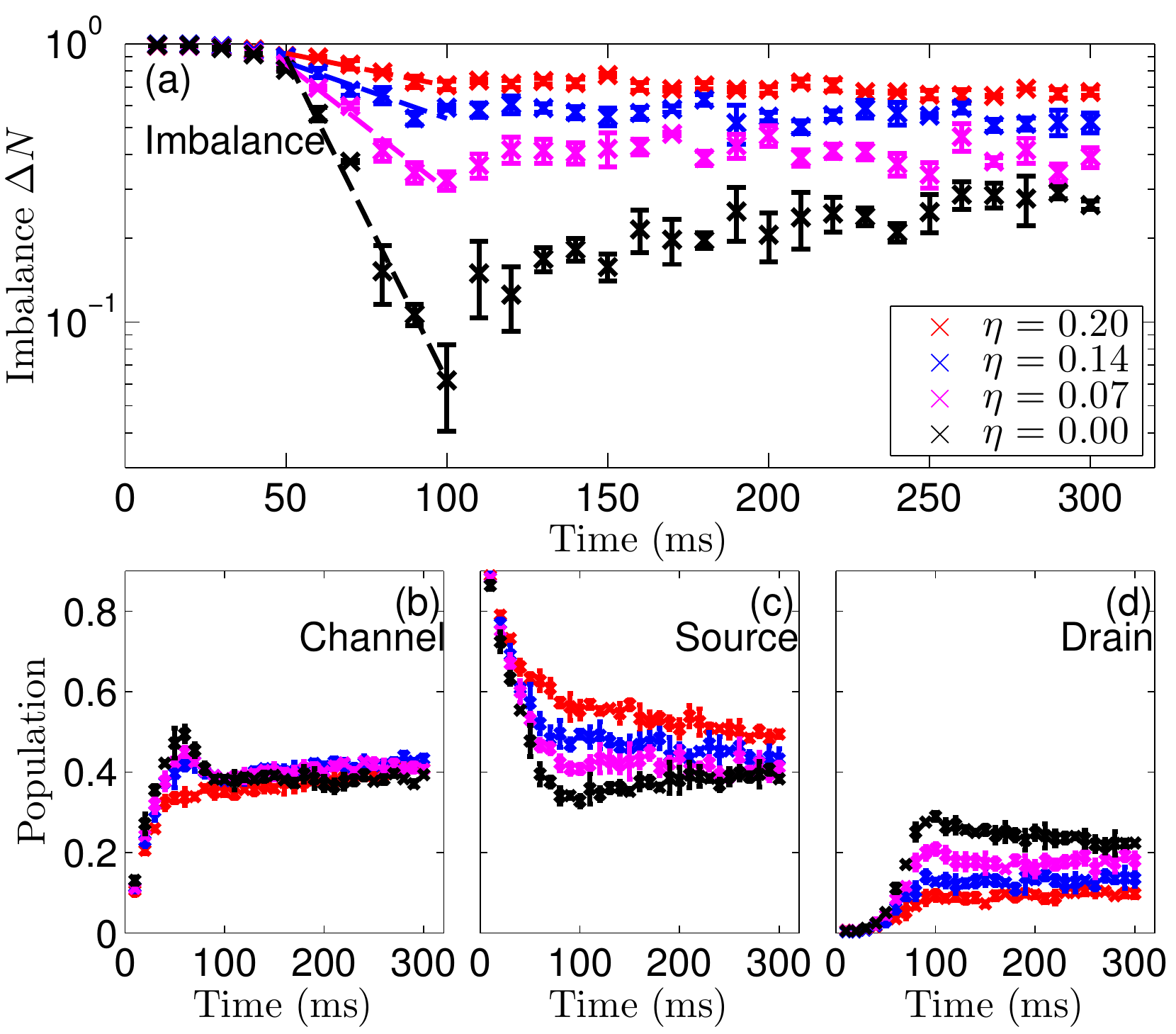}
    \caption{\label{fig:Imbalance}
             \textbf{Temporal evolution of atom populations.}
             (a) The number imbalance $\Delta N$ vs time for four different
             fill-factors, with $(r, L, w) = (43, 162, 36)\,\mu$m. Plots are
             overlaid with the linear fits to the semilogarithmic plot used to
             calculate the resistance via Eq.~(\ref{eq:atomtronic}).
             (b) Evolution of the channel population.
             (c) Evolution of the source reservoir population.
             (d) Evolution of the drain reservoir population. The errorbars show standard deviations in the data, over 3 disorder realisations.
            }
\end{figure}

\section{Results}
\subsection{Theory}
Our experimental observations are complemented with a systematic numerical
analysis in order to understand the experimental findings in more detail and to
support their interpretation. On a fundamental level Anderson localisation is a
single-particle phenomenon, therefore, its appearance in a quantum system can be
captured by a one-body Schr{\"o}dinger equation with a potential term,
$V_{\text{trap}}(\mathbf{r})$, corresponding to the confinement and to the 2D
static disorder. However, in the experiment some weak interaction,
$V_{\text{int}}(\mathbf{r})$, is still present between the particles. The interplay between interactions and localisation is a topic of rigorous debate~\cite{Fishman2012}. We note several theoretical studies suggest that localisation is maintained in the presence of weak interactions in 1D~\cite{Lugan2007b,Lellouch2014,Dujardin2016}, as well as experiments in the many-body localised regime of strong interactions~\cite{Schreiber2015,Lukin2019}. In the presence of interactions the
dynamics are governed by the Gross--Pitaevskii equation \cite{Pethick2002} (GPE)
\begin{equation}
    i \hbar\, \frac{\partial \psi}{\partial t}
    =
    \left \lbrack
        -\frac{\hbar^{2}}{2m} \nabla^2_{\text{2D}} +
        V_{\text{trap}}(\mathbf{r})              +
        V_{\text{int}}(\mathbf{r})
    \right \rbrack
    \psi,
\end{equation}
\noindent which we solve using the adaptive, fourth-order Runge-Kutta-Fehlberg
method~\cite{Burden2011}. Our numerical simulations give access to all
experimentally observed quantities and we present them alongside of the
experimental measurements for comparison. A further advantage of the numerical simulations is that they allow us to switch off the interactions following the initial expansion, allowing us to analyse the effect of interactions on Anderson localisation~\cite{SupplementaryMaterial}.

\subsection{Evolution of Channel Density Profiles}
The experimental setup consists of ultracold atoms propagating from a source reservoir, through a disordered channel, and into a drain reservoir. The optical  setup is illustrated in Fig.~\ref{Fig:ExperimentalSetup}(a), the dumbbell-shaped architecture of the environment is illustrated in Fig.~\ref{Fig:ExperimentalSetup}(b), and the setup is described in detail in Haase et al.~\cite{Haase2017}. We quantify the transport properties of this system in two ways. First, we analyse the long-time
behaviour of the atomic density profile within the channel, which allows direct
observation of exponential localisation. Second, we measure the
temporal behaviour of the source, channel and drain populations ($N_{\mathrm{s}}$, $N_{\mathrm{c}}$,
$N_{\mathrm{d}}$). This facilitates the measurement of the transmission coefficient of the
channel, which we interpret as a channel `resistance' \cite{Eckel2016}.

We first analyse the long-time behaviour of the system. The signature of
Anderson localisation is an exponentially decaying wavefunction, such that the
density decays in space with a localisation length of $\xi$, as:
\begin{equation}
    \rho(x) = \rho_{0} e^{-2 x/\xi}.
\end{equation}
\noindent After many scattering events, the density of atoms within the disordered channel evolves to exhibit an
exponentially decaying profile in an Anderson-localised system. Note that 2D is a special case: although
there is a distribution of atomic momenta, and therefore a
distribution of localisation lengths, the density profile is expected to be
exponential \cite{Miniatura2009}. This is a consequence of the finite (yet
possibly large) localisation length for all momenta in 2D.

Figures~\ref{fig:ComplexFig}(a)-(c) plot the time evolution of the channel density
profile for three different fill-factors. For weak disorder, we observe a near
constant density profile at short evolution times, which evolves to a
non-exponential profile for long times. Highly disordered channels ($\eta\geq
0.17$) show distinctly different behaviour. All evolution times over 50\,ms
indicate an exponential profile. The apparent localisation length, found from
the gradient of $\log(\rho(x))$ curve and plotted in Fig.~\ref{fig:ComplexFig}(d),
approaches a quasi-stationary value for long expansion times. The solution of the Gross--Pitaevskii equation shows similar behaviour, superimposed with an oscillation about a constant value. We extract the localisation length measurement from the mean value for expansion times larger than 200 ms and present the result in Fig.~\ref{fig:ComplexFig}(e). This data indicates that we achieve a localisation length shorter than the channel length of 180\,$\mu$m for $\eta \gtrsim$ 0.25, with a similar threshold observed for the shorter and wider channel meeting the criterion for strong Anderson localisation. We find a clear relationship showing a reduced localisation length with increasing fill-factor. Numerical simulations give localisation lengths in reasonable quantitative agreement with experiment. We also remark on the slightly stronger localisation observed in the wider 58~$\mu$m channel compared to the 43~$\mu$m channel: while a fuller investigation of the width-dependence of the localisation length is planned for further study, here we hypothesise that the longer localisation length in the narrower channel is due to finite size effects, associated with the localisation length being significantly longer than the channel width.

We find a difference in equilibration time between theory and experiment in Fig.~\ref{fig:ComplexFig}(a), though for longer times ($t>400$~ms) we confirm that the numerical simulations do tend to a near-flat constant density profile~\cite{SupplementaryMaterial}.  We attribute differences between the experiment and simulation to effects which are not directly included in the simulation (including finite temperature effects, the smooth disordered potential, and the deviations from flatness in the 2D trap). While these differences may result in minor deviation between experiment and theory, both point towards Anderson localization. We also confirm that the simulations in Figs.~\ref{fig:ComplexFig}(b) and \ref{fig:ComplexFig}(c) exhibit an exponential channel profile with a quasi-stationary mean localisation length for very long times in the case of $\eta\geq 0.17$, for $t > 400$\,ms~\cite{SupplementaryMaterial}.

\subsection{Momentum Dependence}

Can we attribute the observed exponential density profiles to quantum interference (Anderson localisation)? The alternative interpretation would be classical trapping within the disordered potential for atoms with energies below the percolation threshold. The numerical simulations in Fig.~\ref{Fig:MomentumDist}(a) and (b) show the initial $k$-distribution following the interaction driven expansion from the BEC. The plot in Fig.~\ref{Fig:MomentumDist}(e) shows the steady state $k$-distribution within the channel for three different fill-factors. Based on this momentum distribution for $\eta = 0.32$, only 0.8\% of atoms within the channel have an energy below the disorder percolation threshold, as calculated according to Morong and DeMarco~\cite{Morong2015}. This low fraction of classically trapped atoms allows us to be confident that any observed localisation is indeed due to quantum interference. 

The momentum distributions, obtained by numerical simulation and illustrated in Fig.~\ref{Fig:MomentumDist}, provide further insight into the system dynamics. High energy atoms propagate into the drain reservoir. The difference in the drain momentum distributions between zero-disorder and disordered systems (Fig.~\ref{Fig:MomentumDist}(f)) shows that the disordered channel acts as an effective energy-filter, preventing low energy atoms ($|k|\lesssim 1.5~\mu$m$^{-1}$) from propagating into the drain. We interpret the complete inhibition of propagation of low-energy atoms as signifying Anderson localisation. The filtering effect is slightly stronger for $\eta=0.32$ compared to $\eta=0.17$. Weakly localised atoms are in an extended state of the system and are able to accumulate in the drain. The momentum distribution in the source (Fig.~\ref{Fig:MomentumDist}(d)) is skewed to low energy, because the dwell time within the source reservoir is inversely proportional to $|k|$. The channel (Fig.~\ref{Fig:MomentumDist}(e)) contains a mixture of weakly localised high-energy atoms in an extended state across the dumbbell, and strongly localised low-energy atoms. We note that the channel clearly contains a larger number of very low energy atoms ($|k|<1~\mu$m$^{-1}$) when disorder is present, indicating that these low-energy atoms are localised within the channel.

In Fig.~\ref{Fig:MomentumDist}(c), we plot the localisation length expected according to:

\begin{equation}
\xi(|k|)=\ell_{\mathrm{s}} e^{\pi |k|\ell_{\mathrm{tr}}/2},
\label{eq:loclength2d}
\end{equation}

\noindent where $\ell_{\mathrm{s}}\approx\sigma/\sqrt{\eta}$ is the scattering mean free path and $\ell_{\mathrm{tr}}=\Lambda(|k|\sigma)\ell_{\mathrm{s}}$ is the transport mean free path~\cite{Muller2005,Lee1985,SupplementaryMaterial}. The curve in Fig.~\ref{Fig:MomentumDist}(c), based on estimates of the mean free path within the system, predicts localisation lengths which are shorter than the system size for $|k|\lesssim 0.55~\mu$m$^{-1}$. We emphasise that this estimate should be considered in the context of the sensitive exponential dependence of the parameters, the specific microscopic details of the disorder, and the finite size of the system, which are not included in the general estimate of Eq.~(\ref{eq:loclength2d}). We note that previous theoretical investigations using point disorder obtained a sub-exponential dependence of $\xi(|k|)$~\cite{Morong2015}, with localisation lengths on the order of 100~$\mu$m expected up to $|k|=6~\mu$m$^{-1}$, in a system with mean free paths of $\ell_{\mathrm{s}}\approx\ell_{\mathrm{tr}}\approx 2~\mu$m, similar to our experiment. We conclude that our experimental regime is within the bounds set by the established theory, in which Anderson localisation can be expected to be observed. At the same time, classical trapping plays a negligible role in the dynamics within the channel.

\subsection{Effect of Interactions}
The role which interactions play in Anderson localisation has been richly debated in the literature~\cite{Lee1985,Shepelyansky1993,Fishman2012,Schreiber2015,Dujardin2016,Donsa2017}. This experiment is conducted with ~$1.6\times 10^4$ atoms, resulting in an average density of $\sim$1 atom / $\mu$m$^2$. With this level of atomic density, the interaction energy is significantly lower than either the mean kinetic energy or the disordered potential depth. The experiment is conducted in a regime of weak repulsive interaction, and our numerical simulations indicate that the observed localisation length would be unchanged within error for the non-interacting case. Based on our numerical analysis, we estimate that interaction strengths more than 5 times the experimental interaction would be required to significantly alter the observed density profiles~\cite{SupplementaryMaterial}.

\subsection{Atomtronic Analysis}
For a second complementary analysis, we treat the system as an `atomtronic'
circuit \cite{Eckel2016} and describe the transport in terms of the atomic
current flowing between two reservoirs of capacitance $C$ but impeded by a
channel resistance $R$. The atomic current is defined by the number imbalance
between the source and drain reservoirs: $\Delta N = (N_{\mathrm{s}} - N_{\mathrm{d}})/(N_{\mathrm{d}} + N_{\mathrm{s}})$.
Esslinger and co-workers suggested \cite{Brantut2012} the phenomenological
relation
\begin{equation}
    \label{eq:atomtronic}
    \frac{d\Delta N}{dt} = -\frac{\Delta N}{RC}.
\end{equation}

\noindent The data in Fig.~\ref{fig:Imbalance}a shows the evolution of $\Delta N$ for
varying fill-factors and for three timescales. In the ballistic period atoms
transport across the channel and arrive at the second reservoir. In this first
period, the imbalance remains unity due to an empty second reservoir. Following
the ballistic time, there is a period of $\sim 40$\,ms during which the imbalance
reduces at its greatest rate. This initial transfer rate is greatest for zero
disorder. In this period, we find an approximately linear relation between
$\log(\Delta N)$ and time, supporting the RC circuit model \eqref{eq:atomtronic},
and we use this transport period to measure the channel resistance. Finally, the
system moves into a third regime of transport, in which the number imbalance
approaches a steady-state, non-zero value for finite $\eta$. This steady-state
behaviour is 
supported by our numerical GPE simulations. We interpret this non-zero steady
state number imbalance to be a consequence of a combination of Anderson
localisation, and enhanced reflection into the source reservoir due to weak localisation. We note that disorder with low fill-factor (e.g., $\eta
=0.07$) significantly reduces transport, as evidenced by the non-zero
steady-state imbalance, although exponential localisation is not observed in
this case (cf. Fig.~\ref{fig:ComplexFig}a). The reduction in transport is a
significant observation, due to the near-zero percolation threshold of $\eta =
0.07$ disorder \cite{Morong2015}.%

In Fig.~\ref{fig:Imbalance}(b)-(d) we show the populations of the channel,
source and drain reservoirs as a function of time. While the reservoir
populations show a dependence on $\eta$, the channel population is largely independent of $\eta$ and approaches a steady state. Weak localisation effects within the channel lead to an enhanced reflection coefficient into the source reservoir, and we estimate the mean dwell time within the channel to be 110~ms~\cite{Pierrat2014}. This estimated dwell time is largely independent of the details of the disorder and coincides with the population equilibration time. We also note that the steady-state channel population agrees with the relative area of the channel with respect to the whole dumbbell.

\begin{figure}[ht]
    \centering
    \includegraphics[width=250px]{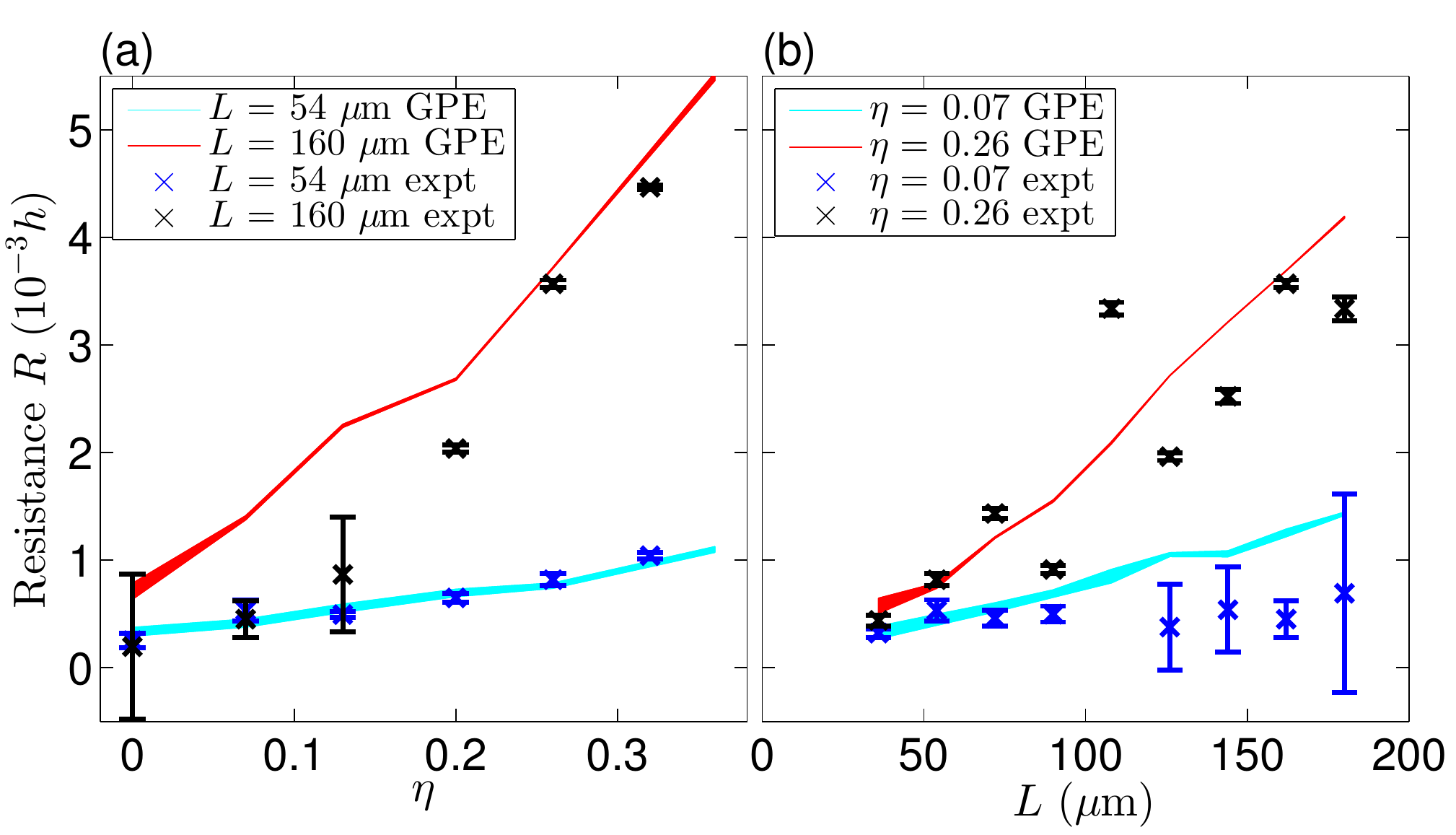}
    \caption{\label{fig:Resistance_vs_FillFactor}
             \textbf{Channel resistance measurement.}
             (a) The resistance as a function of fill-factor, for two channel
             lengths, in units of Planck's constant, with $(r,w)=(43,36)~\mu$m.
             (b) The resistance a function of length, for two fill-factors, with $(r,w)=(43,36)~\mu$m.
             Results are overlaid with GPE simulations, with the shaded region
             indicating one standard-deviation on the simulation value. The errorbars show standard deviations in the data, over 3 disorder realisations.}
\end{figure}
The resistance in units of Planck's constant, $h$, is plotted in
Fig.~\ref{fig:Resistance_vs_FillFactor} for a range of fill-factors and lengths. In this system, $\hbar C = 19$~s~\cite{SupplementaryMaterial}. We observe a stronger dependence of the resistance on fill-factor for longer
channel lengths; likewise, we observe a stronger dependence of the resistance on
channel length for stronger disorder. While we expect the resistance to be
exponential in the channel length in the strongly localised regime
\cite{Abrahams1979}, here we observe a slower dependence within the accessible
experimental parameters. Figure~\ref{Fig:MomentumDist}(f) shows that the atoms in the drain have significantly higher energy than the atoms in the channel or source, and we conclude that the main contribution to the
resistance measurement comes from atoms with very large localisation lengths and energies larger than the mean energy. We note
close agreement between the experimental data and numerical simulations for the
resistance measurements.

\section{Discussion}
In conclusion, in combining a highly tuneable experimental platform with full numerical GPE simulations, we have provided compelling evidence for Anderson localisation in a two-dimensional ultracold atom system. For atoms traversing a disordered 2D potential of point scatterers in a regime of weak atomic interaction, we demonstrate clear signatures of exponential localisation. We observe temporally-stable exponential channel profiles for $\eta\geq 0.17$. The logarithm of these  profiles are linear and do not change significantly for $t>100$\,ms. We have shown for our system that this localisation cannot be explained by classical trapping within the channel. The supporting numerical simulations show that transport of low energy atoms is almost totally inhibited by the disordered channel. We therefore interpret profiles with localisation lengths shorter than the channel length to signify Anderson localisation in 2D.

Through measurements of the localisation length, we have demonstrated that the transport may be tuned via the disorder fill-factor from a regime of ballistic, to diffusive, and then to strongly localised transport with $\xi<L$. The dumbbell-shaped architecture enabled two complementary analyses, allowing measurements of the
channel resistance, together with the in-channel density evolution. The channel
resistance indicates that while atoms traversing weak disorder ($\eta=0.07$) do
not exhibit Anderson localisation, the transport is significantly reduced from
the zero-disorder case, despite the near-zero percolation threshold. All
experimental observations are supported by zero-temperature
Gross-Pitaevskii calculations, and the experimental conditions are within the bounds for observation of localisation set by the established theory. The numerical simulations reproduce all signatures observed in the experiment, differing only in equilibration time. The simulations provide additional insight into the role of interactions and the momentum distributions at different fill factors, corroborating the experimental evidence, and providing strong support that Anderson localisation is the suitable interpretation of the exponential density profiles and of the reduced transport. These results provide a springboard for studying localisation and the causes of delocalisation in 2D systems with a quantum-simulator-like device.

%

\subsection*{Acknowledgements}%
    The authors would like to thank A.~V.~H.~McPhail and I. Herrera for
    laboratory assistance, and S.~S.~Shamailov for detailed discussions. D.~H.~W. thanks L. Sanchez-Palencia and D. Delande
    for discussions. C.~G would like to thank the German Academic Exchange
    Service (DAAD) for financial support during his stay at the University of
    Otago. This work was supported by the Marsden Fund, grant number UOA1330,
    administered by the Royal Society of New Zealand.

\subsection*{Author Contributions}
    M.~D.~H. and D.~H.~W. planned the research. T.~A.~H., D.~H.~W., and D.~J.~B.
    constructed the experiment. T.~A.~H. performed the measurements, with
    D.~J.~B. and D.~H.~W. providing assistance. D.~J.~B. and D.~H.~W. carried out the data
    analysis. J.~H. wrote the Gross-Pitaevskii code. M.~S.~N. and D.~S ran and
    analysed the simulations, and together with C.~G and
    D.~A.~W.~H. formed the theoretical underpinning. M.~D.~H. and D.~A.~W.~H.
    supervised the experimental and theoretical work, respectively. All authors
    discussed the research and contributed to the manuscript.

\subsection*{Competing Interests}
    The authors declare no competing interests.

\subsection*{Data availability.}
    All data presented in this publication is available upon request.

\subsection*{Correspondence}
    Correspondence and requests for materials related to the experiment can be
    addressed to Maarten D.~Hoogerland (\texttt{m.hoogerland@auckland.ac.nz}),
    while queries regarding the theoretical investigation may be directed to
    David A.~W.~Hutchinson (\texttt{david.hutchinson@otago.ac.nz}).


\end{document}


\title{Supplementary Material --- Observation of two-dimensional Anderson localisation of ultracold atoms}

\author{Donald H. White}
    \altaffiliation[Present address:\ ]{Department of Applied Physics, Waseda University, Shinjuku, Tokyo, Japan}
    \affiliation{Dodd-Walls Centre for Photonic and Quantum Technologies, New Zealand}
    \affiliation{Department of Physics, University of Auckland, Auckland, New Zealand}
\author{Thomas A.~Haase}
    \thanks{D.\,H.\,W. and T.\,A.\,H. contributed equally to this work.}
    \affiliation{Dodd-Walls Centre for Photonic and Quantum Technologies, New Zealand}
    \affiliation{Department of Physics, University of Auckland, Auckland, New Zealand}
\author{Dylan J.~Brown}
    \affiliation{Dodd-Walls Centre for Photonic and Quantum Technologies, New Zealand}
    \affiliation{Department of Physics, University of Auckland, Auckland, New Zealand}
\author{Maarten D.~Hoogerland}
    \affiliation{Dodd-Walls Centre for Photonic and Quantum Technologies, New Zealand}
    \affiliation{Department of Physics, University of Auckland, Auckland, New Zealand}

\author{\\ Mojdeh S.~Najafabadi}
    \affiliation{Dodd-Walls Centre for Photonic and Quantum Technologies, New Zealand}
    \affiliation{Department of Physics, University of Otago, Dunedin, New Zealand}
\author{John L.~Helm}
    \affiliation{Dodd-Walls Centre for Photonic and Quantum Technologies, New Zealand}
    \affiliation{Department of Physics, University of Otago, Dunedin, New Zealand}
\author{Christopher Gies}
    \affiliation{Institut f{\"u}r Theoretische Physik, Universit{\"a}t Bremen, Bremen, Germany}
\author{Daniel Schumayer}
    \affiliation{Dodd-Walls Centre for Photonic and Quantum Technologies, New Zealand}
    \affiliation{Department of Physics, University of Otago, Dunedin, New Zealand}
\author{David A.~W.~Hutchinson}
    \affiliation{Dodd-Walls Centre for Photonic and Quantum Technologies, New Zealand}
    \affiliation{Department of Physics, University of Otago, Dunedin, New Zealand}

\maketitle

\textbf{
Here we provide additional supporting information for the
results described in the main text ``Observation of two-dimensional Anderson localisation of ultracold atoms''.}

\section{Experimental details}

This experiment has been designed to be a `quantum simulator' of 2D transport
physics. This requires full knowledge of the topography of the 2D potential
landscape. Ideally, the basic 2D trap would be flat, with disorder and
boundaries introduced by the SLM-projected landscape. While we approximate this
condition, we have found the presence of `fringes' within the 2D trap. These
fringes run along the $y$-direction, and have a period of $\approx 150~\mu$m.
The fringe depth is on the order of 5~nK. We attribute the fringes to the
interference within the 1064~nm beams, which occurs from distortion to the
phase-front of the beam from the vacuum window. The fringe central position may
be adjusted by relative horizontal alignment of the two interfering 1064~nm
beams, and we set the fringe centre to overlap the centre of the source
reservoir.

The BEC is initially formed in a CO$_2$ crossed beam laser trap, and adiabatically loaded into the two-dimensional trap by ramping down the CO$_2$ laser power. Prior to the release of the CO$_2$ laser trap, the trap frequency in the horizontal directions is on the order of $\omega_{\mathrm{x,y}}\approx 2\pi\cdot 50$~Hz. The dimensionless interaction strength $\tilde{g} = a_{\mathrm{s}}\sqrt{8\pi m\omega_{\mathrm{z}}/\hbar} = 0.07$ . The peak density within the CO$_2$ laser trap is $n_0\approx \frac{m}{\hbar}\sqrt{\frac{N}{\pi\tilde{g}}}\omega_{\mathrm{x,y}}\approx 120$ atoms/$\mu\text{m}^2$. The healing length $\xi=1/\sqrt{n_0\tilde{g}}=350$~nm. The chemical potential within the CO$_2$ laser trap is $\mu =\hbar^2 n_0\tilde{g}/m$, and $\mu/k_{\mathrm{B}}=40$~nK. Once the atoms have been released from the CO$_2$ laser trap and expanded into the dumbbell, the density reduces to the order of 1 atom$/\mu\text{m}^2$, reducing the chemical potential to $\mu/k_{\mathrm{B}}=0.3$~nK~\cite{Petrov2000,Kruger2007}.

A linear slope, with acceleration 0.002~m/s, is applied to the 2D trap, meaning
that atoms are no longer bound by the fringe, and may enter the trap. While the
presence of the fringe affects the bulk motion of the atoms within the channel,
the period of the fringe is larger than the channel lengths used, and this means
that the Anderson localisation physics should be unchanged. We note close
agreement with GPE simulations (discussed below), even if the fringe pattern is
excluded.

The BEC is placed such that its centre lies about half way between the centre of
the source reservoir and the channel opening. We calibrate the acceleration by
adjusting the tilt such that the first wave of atoms arrives at the far end of
the drain reservoir at 100 ms after releasing the atoms from the dipole trap,
using a 144 $\mu$m long channel.


\section{Full datasets}
\begin{figure}[bth]
    \includegraphics[width=250px]{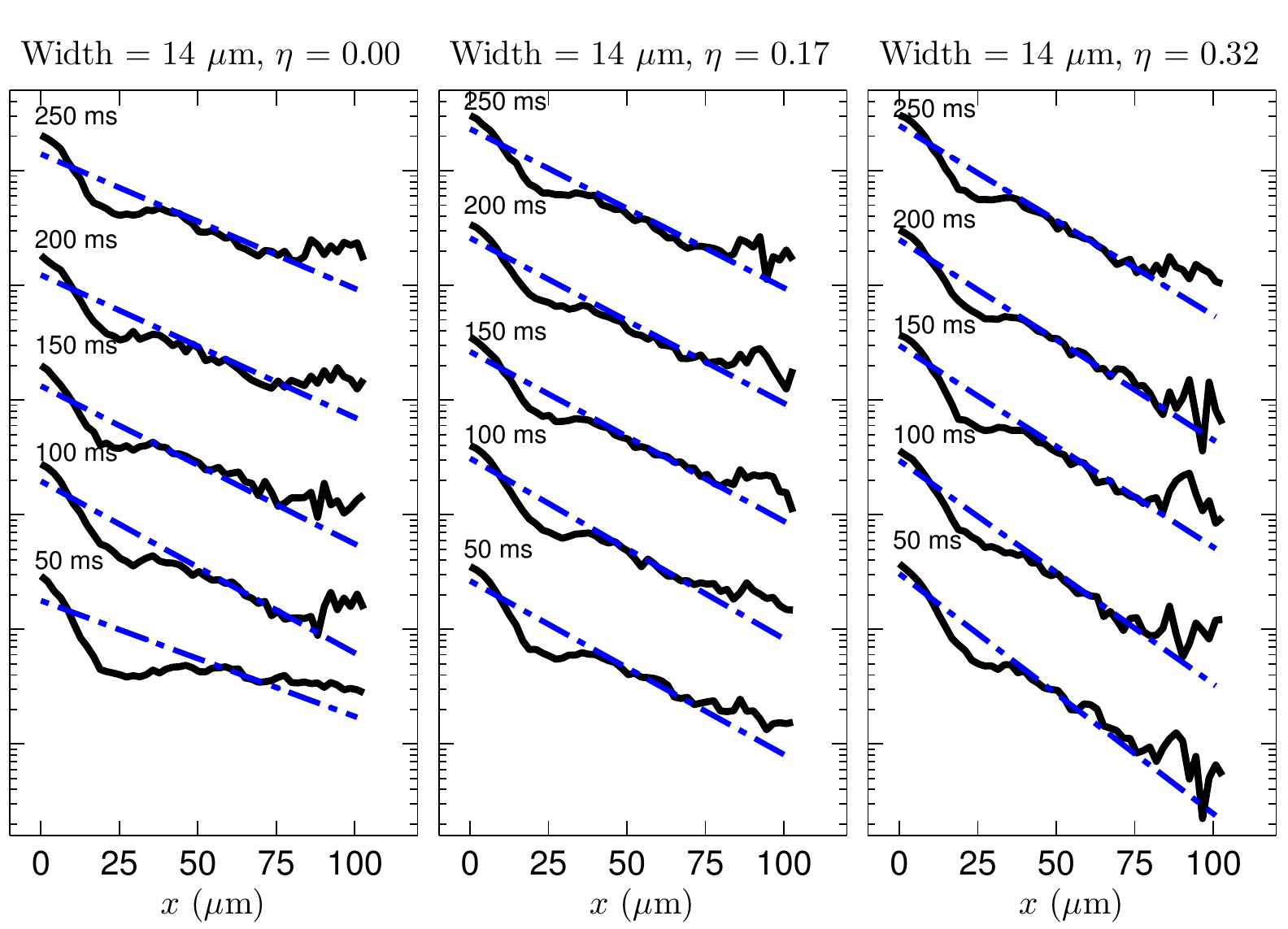}
    \caption{\label{fig:wid20}
             \textbf{14~$\mu$m width channel density profile data.} Channel profiles from 50~ms to 250~ms are plotted on a semilogarithmic scale, together with a linear fit to the logarithmic data. This data is an average of three experimental disorder realisations.}

\end{figure}

\begin{figure}[bth]
    \includegraphics[width=250px]{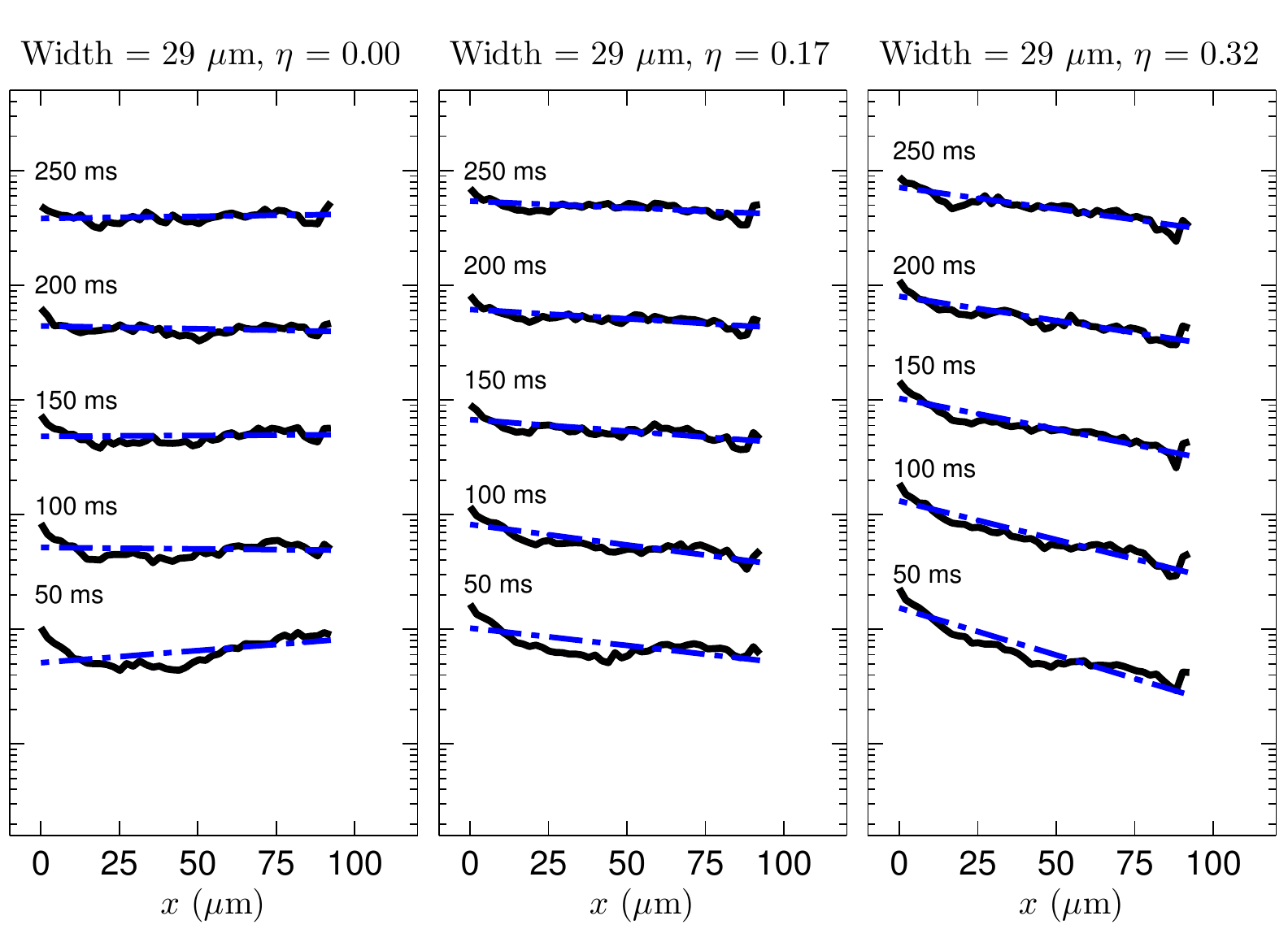}
    \caption{\label{fig:wid40}
             \textbf{29~$\mu$m width channel density profile data.} Channel profiles from 50~ms to 250~ms are plotted on a semilogarithmic scale, together with a linear fit to the logarithmic data. This data is an average of three experimental disorder realisations.}

\end{figure}

\begin{figure}[bth]
    \includegraphics[width=250px]{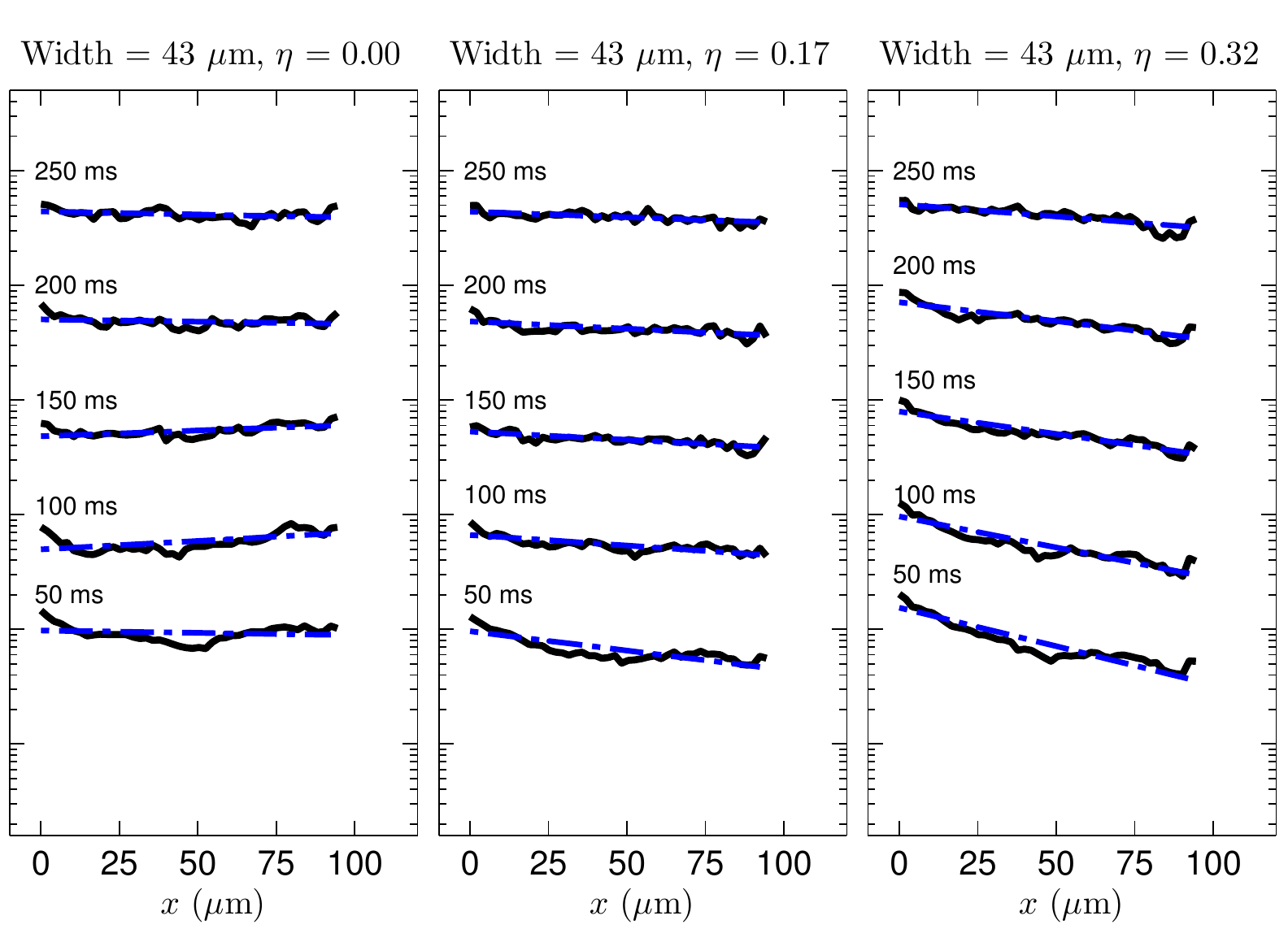}
    \caption{\label{fig:wid60}
             \textbf{43~$\mu$m width channel density profile data.} Channel profiles from 50~ms to 250~ms are plotted on a semilogarithmic scale, together with a linear fit to the logarithmic data. This data is an average of three experimental disorder realisations.}

\end{figure}

\begin{figure}[bth]
    \includegraphics[width=250px]{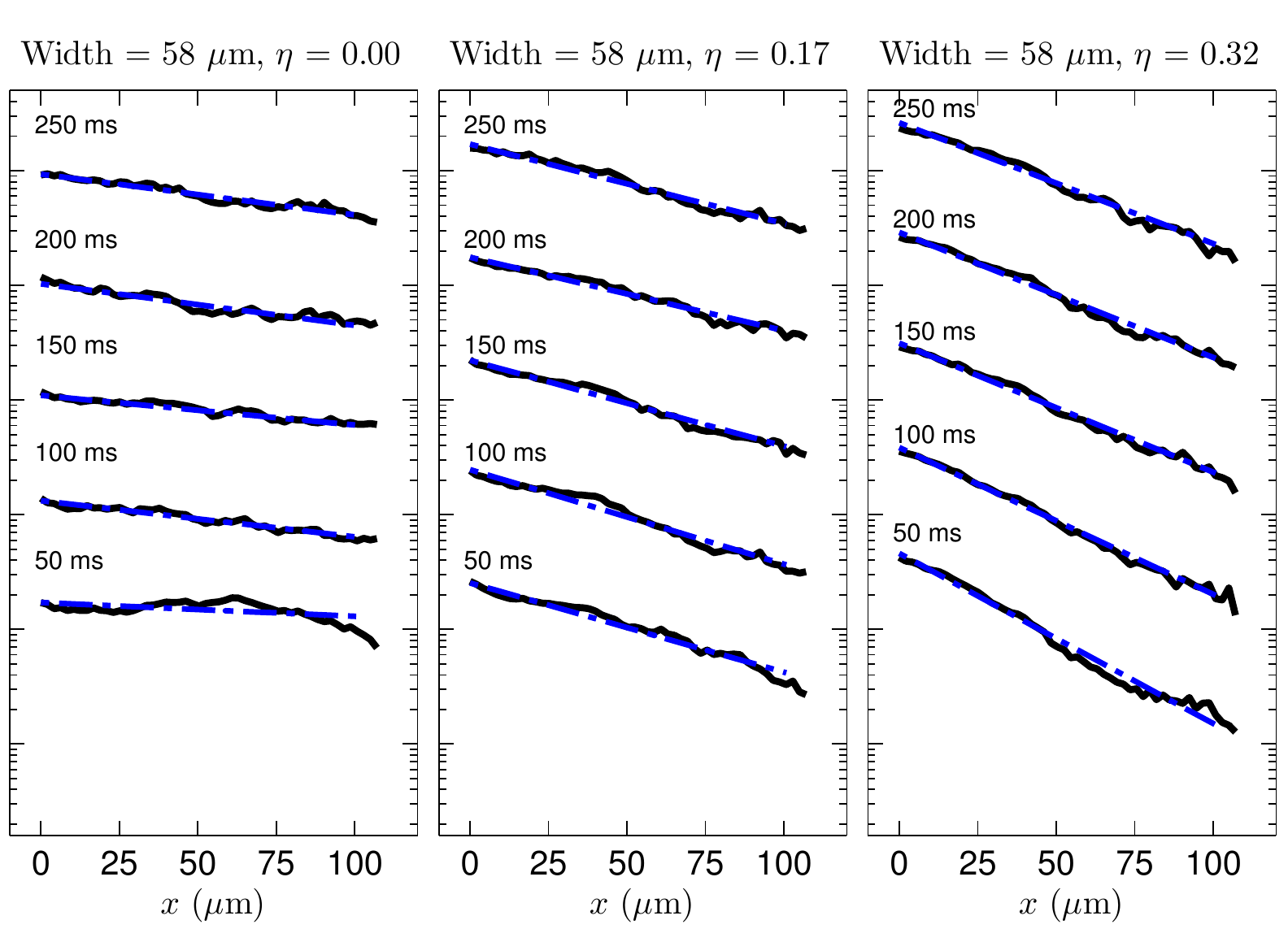}
    \caption{\label{fig:wid80}
             \textbf{58~$\mu$m width channel density profile data.} Channel profiles from 50~ms to 250~ms are plotted on a semilogarithmic scale, together with a linear fit to the logarithmic data. This data is an average of three experimental disorder realisations.}

\end{figure}

\begin{figure}[bth]
    \includegraphics[width=250px]{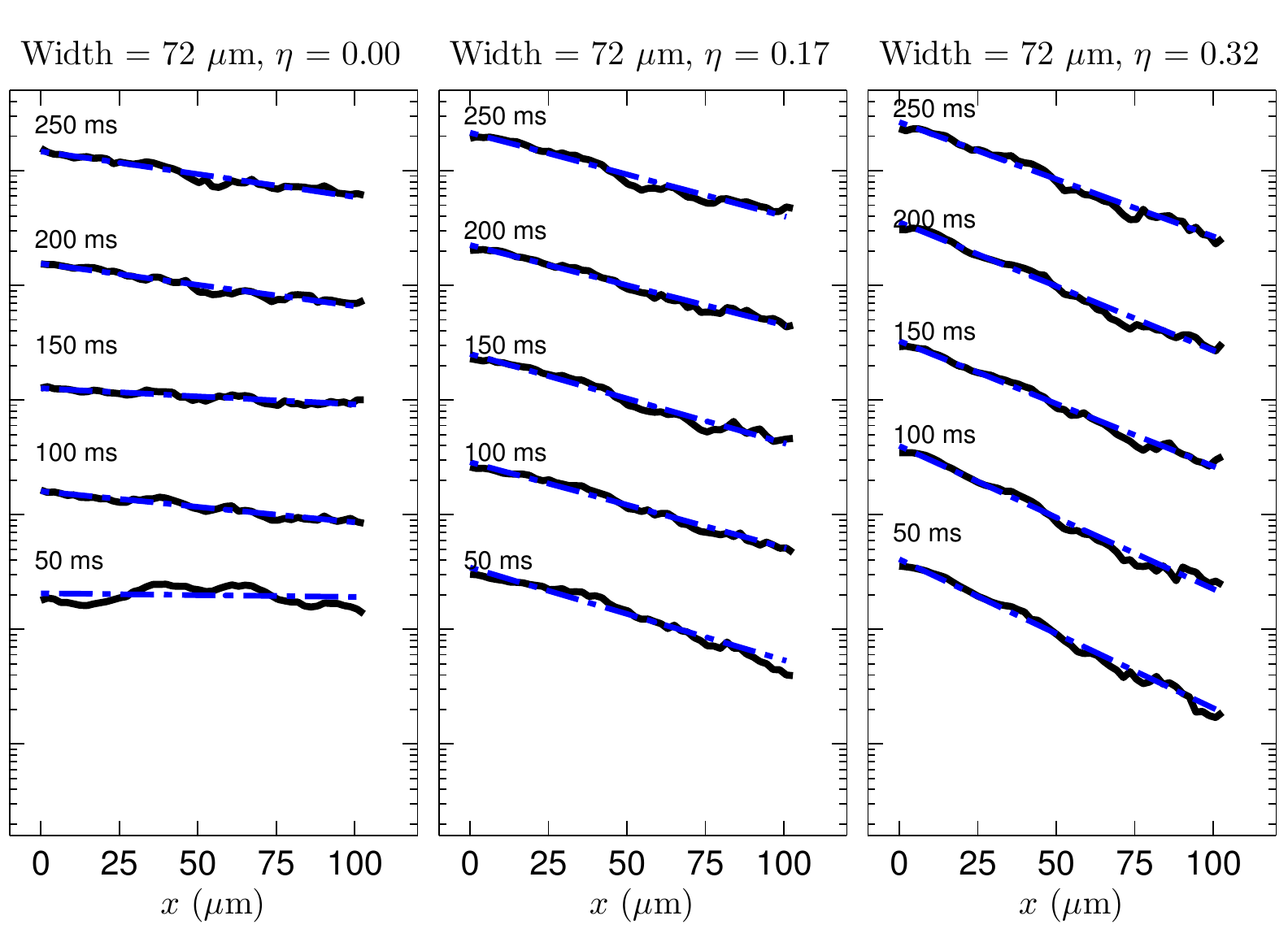}
    \caption{\label{fig:wid100}
             \textbf{72~$\mu$m width channel density profile data.} Channel profiles from 50~ms to 250~ms are plotted on a semilogarithmic scale, together with a linear fit to the logarithmic data. This data is an average of three experimental disorder realisations.}

\end{figure}

\begin{figure}[bth]
    \includegraphics[width=250px]{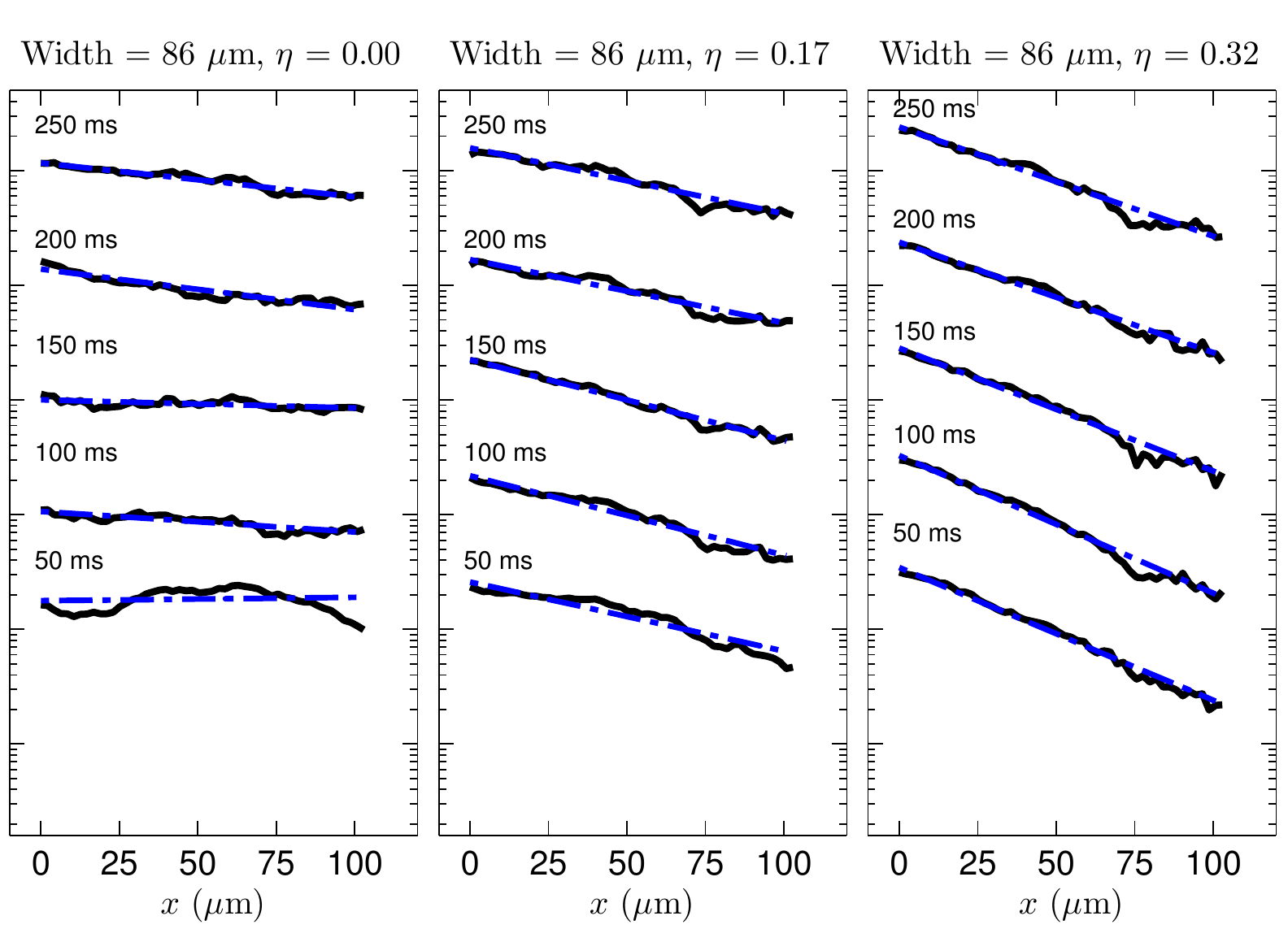}
    \caption{\label{fig:wid120}
             \textbf{86~$\mu$m width channel density profile data.} Channel profiles from 50~ms to 250~ms are plotted on a semilogarithmic scale, together with a linear fit to the logarithmic data. This data is an average of three experimental disorder realisations.}

\end{figure}

We include in Figs.~\ref{fig:wid20}--\ref{fig:wid120} the full datasets for a range of channel widths. The data indicates that steady-state exponentially-localised channel profiles are obtained for a broad range of channel widths in the presence of disorder. The exception to this is the 14~$\mu$m width channel, where the profile is non-exponential, and exhibits a non-uniform channel profile in the case of zero applied disorder. In this case, we are in a nearly one-dimensional regime, and minor channel disorder due to imperfections in the flat disordered potential results in localised eigenfunctions. In addition, there is significant reflection at the mouth of the channel due to the small channel opening, resulting in the dropoff in channel density seen at the channel opening.


\section{Width dependence}

\begin{figure}[bth]
    \includegraphics[width=250px]{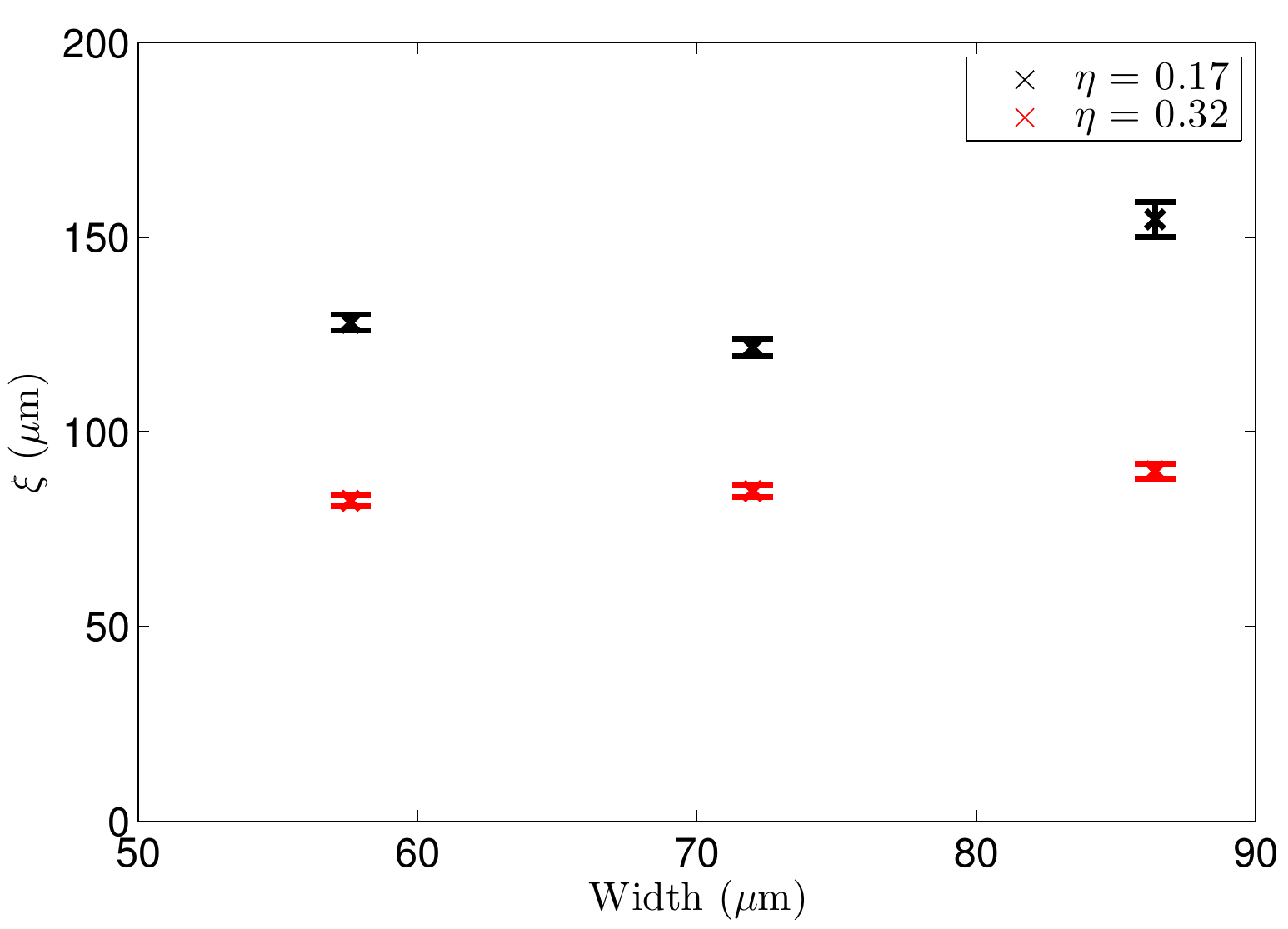}
    \caption{\label{fig:width}
             \textbf{Dependence of localisation length on width for a quasi-2D environment.} The experimental localisation length as an average from 210-250 ms of expansion time is plotted for the width data shown in Figs.~\ref{fig:wid80}-\ref{fig:wid120}, for $\eta=0.17$ and $\eta=0.32$, with $L=108~\mu$m. The error bars show the standard error in the mean over the three disorder configurations.}

\end{figure}

Figure~\ref{fig:width} shows the localisation lengths extracted from the density profiles in Figs.~\ref{fig:wid80}-\ref{fig:wid120}. For $w\lesssim 50~\mu$m, the localisation length is significantly larger than the channel width and finite size effects are strong. For the quasi-2D environment of $w\gtrsim 50~\mu$m, in which the localisation length is of the same order as the channel width, we show in Fig.~\ref{fig:width} that the width does not have a significant influence on the observed localisation length.

\section{Simulations for long times}

\begin{figure}[bth]
    \includegraphics[width=250px]{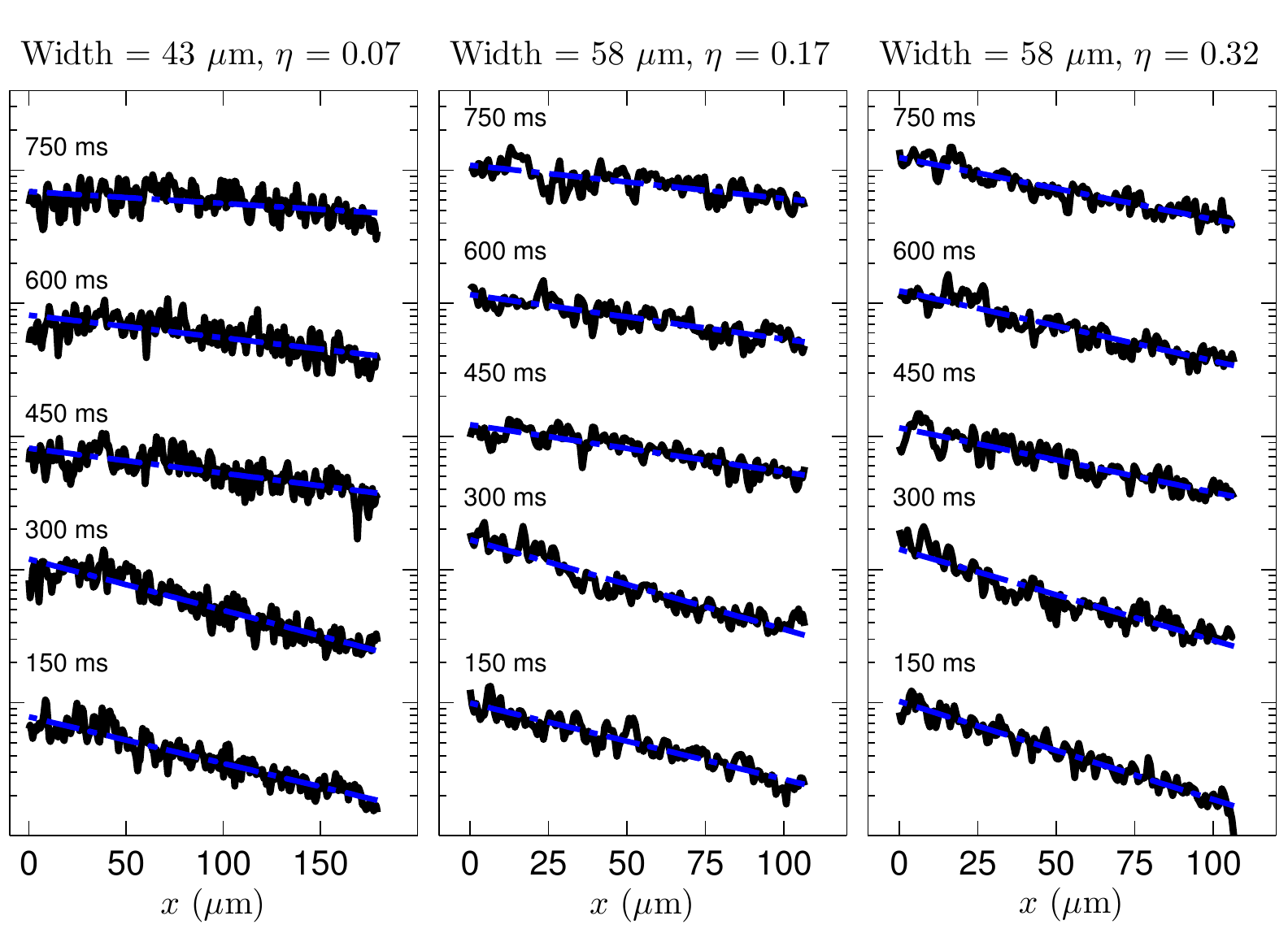}
    \caption{\label{fig:suppl_longtime}
             \textbf{Long-time simulations.} Simulations for the conditions in Fig.~2 of the main manuscript, plotted semilogarithmically for long times. The simulations are overlaid with exponential fits to the data. Offsets by increasing factors of 10 are for clarity.}
\end{figure}

We show in Fig.~\ref{fig:suppl_longtime} simulations conducted for long times for the same conditions as Fig.~2 of the main manuscript. The data show that in the case of $\eta=0.07$, a long-time exponential density profile is not obtained, but the density profile remains exponential for $\eta=0.17$ and $\eta=0.32$. We note that this exponential character remains, despite complications from higher energy atoms re-entering the channel from the drain reservoir.

\section{Momentum distributions}

\begin{figure}[bth]
    \includegraphics[width=250px]{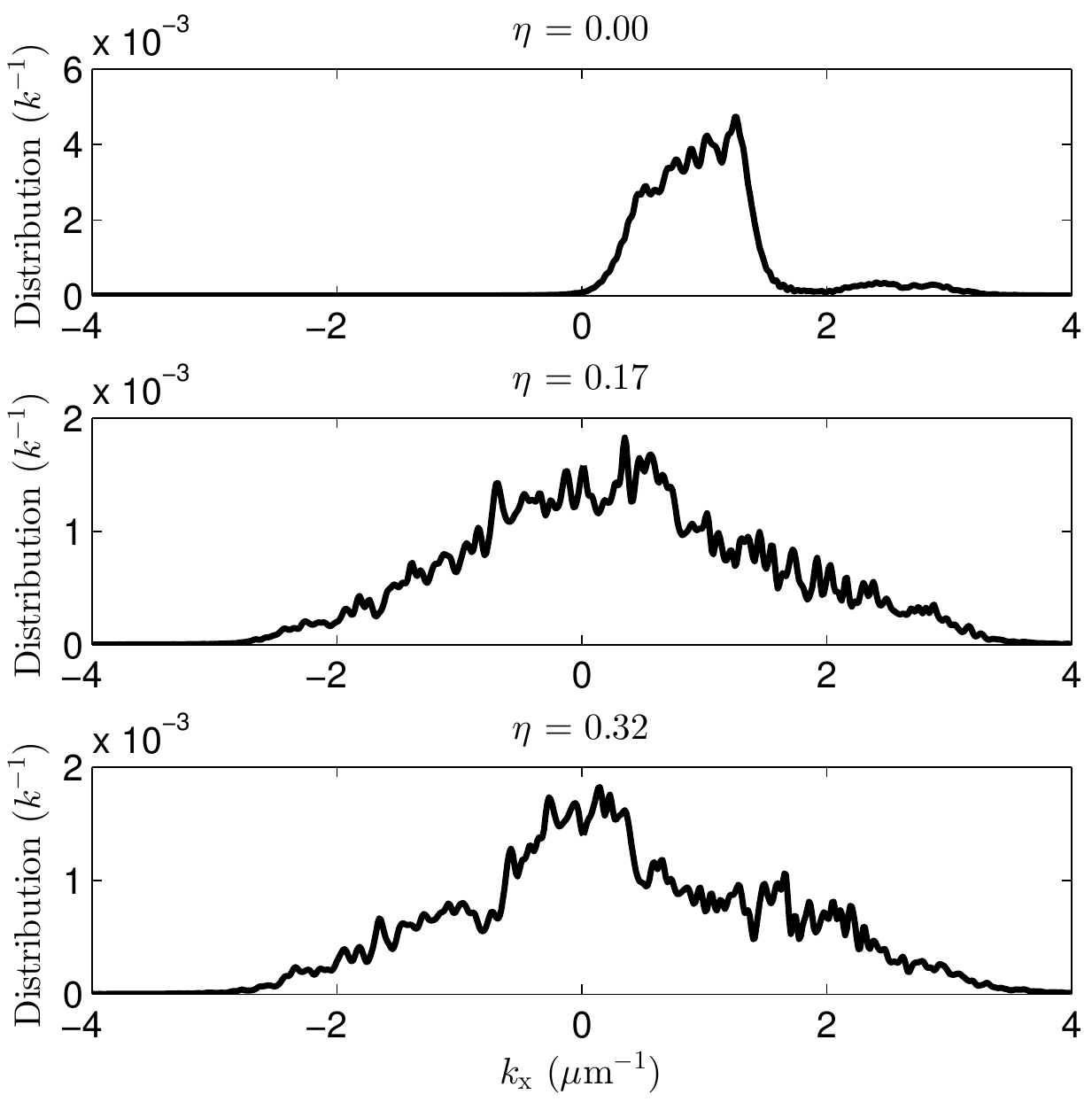}
    \caption{\label{fig:suppl_kx}
             \textbf{Channel momentum distributions.} The momentum distributions along $x$ within the channel are obtained after 125 ms of expansion for three different fill-factors, with ($r,L,w$)=(43,108,58)~$\mu$m, via numerical simulation.}
\end{figure}

We extend the momentum distributions plotted in Fig.~3 of the main manuscript in Fig.~\ref{fig:suppl_kx}, by including the momentum distributions along the $x$-axis within the channel, derived from numerical simulation. (The $x$-axis is the direction along the channel, and a positive value means that $k_{\mathrm{x}}$ is directed towards the drain). This data also gives information regarding the direction of wave propagation, and allows us to draw conclusions regarding the scattering. The data collection time of 125 ms is chosen so as to be prior to atoms from the drain reflecting and re-entering the channel.

In the case of zero disorder, only atoms with a positive $k_{\mathrm{x}}$ value are present within the channel, as can be readily understood from the system geometry. In the case of disorder, scattering alters the direction of $k$, which smooths the $k_{\mathrm{x}}$ distribution. Moreover, we observe a bimodal distribution, consisting of a narrow distribution centred at $k_{\mathrm{x}}=0$, on top of a far broader distribution. This is further evidence that the channel contains atoms with a range of energies, with atoms in the narrow low-energy distribution experiencing exponential localisation and those in the higher energy distribution subject to weak localisation.

\section{Potential slices}

\begin{figure}[bth]
    \includegraphics[width=250px]{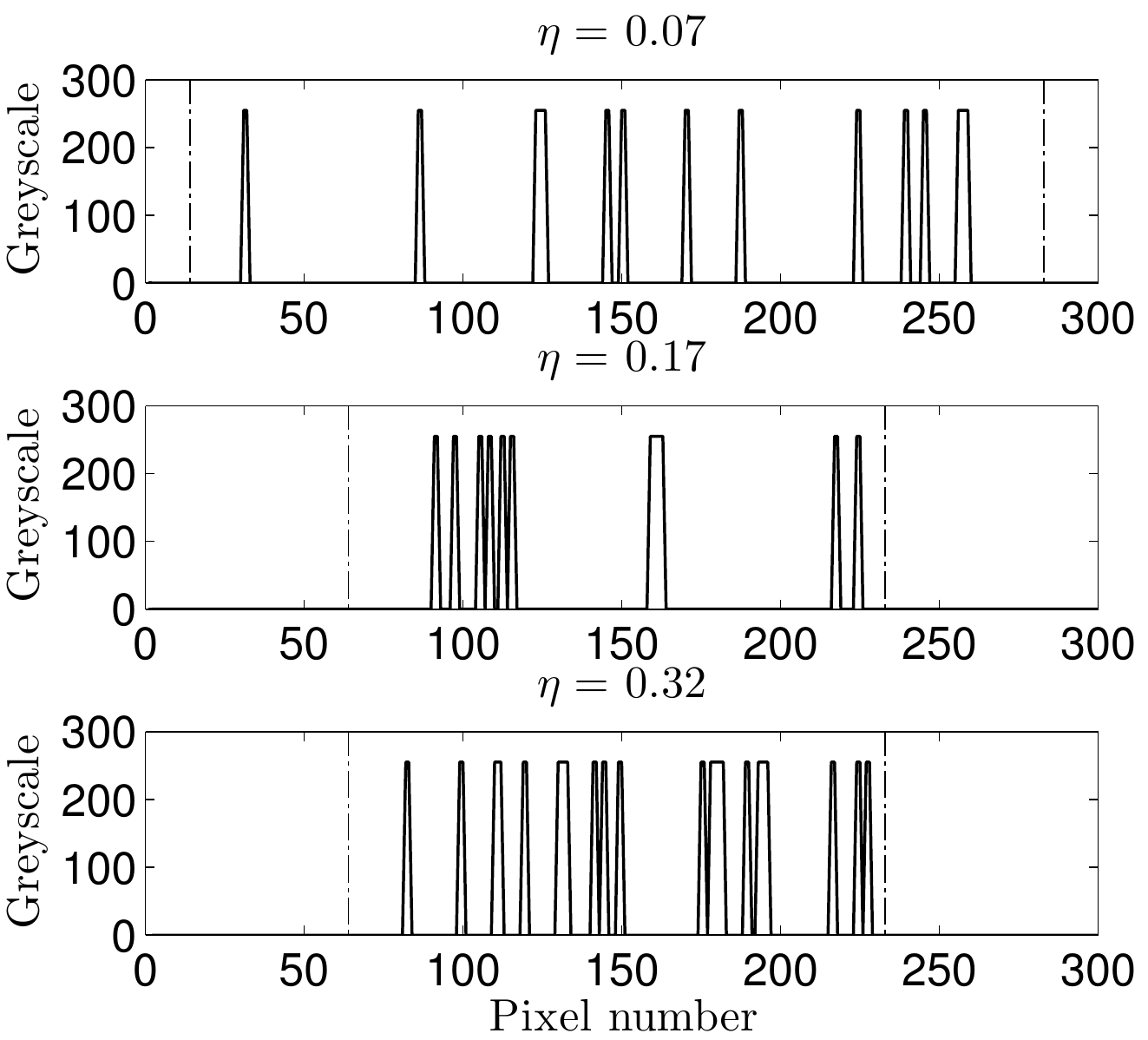}
    \caption{\label{fig:potCut}
             \textbf{Slices of dumbbell potential.} Single slices of the potentials plotted in Fig.~2(a)-(c) of the main text are shown here, along the $x$-direction. Dash-dot lines indicate the mouth of the channel.}
\end{figure}

Figure~\ref{fig:potCut} shows single slices of the potential applied to the spatial light modulator, as used in the experiment for three fill-factors. Each point scatterer occupies 2 pixels along the $x$-direction. Equivalent potentials are used for the numerical simulation. On average, a single slice will yield a fraction $\eta$ of pixels within the channel of greyscale 255.

\section{Reservoir capacitance}

The channel resistance is found from the initial flow of atoms into the drain
reservoir, according to
\begin{equation*}
    \frac{d\Delta N}{dt}
    =
    -\frac{\Delta N}{RC}
\end{equation*}
with the reservoir capacitance given by~\cite{Li2016}
\begin{equation*}
    \label{eq:defineCapacitance}
    C
    =
    \frac{3\left(\frac{1}{2}N\right)^{1/3}}{4\alpha},
\end{equation*}
the constant $\alpha$ is
\begin{equation*}
    \label{eq:slm_defineAlpha}
    \alpha
    =
    \left \lbrack
        \frac{g \, (\frac{1}{2} m \omega_{\mathrm{z}}^2)^{1/2}}
             {\frac{4}{3}\pi r^2}
    \right \rbrack^{2/3},
\end{equation*}
and $g = 4\pi\hbar^2 a_{\mathrm{s}}/m$ is the 3D nonlinearity, $r$ is the reservoir
radius, $m$ is the mass of an atom, $N$ is the number of atoms, and $\omega_{\mathrm{z}}$
is the vertical trapping frequency. In our system with $r=43\,\mu$m, $\hbar
C=19$\,s.

\section{Comparison between ordered and disordered scatterers}

\begin{figure}
    \includegraphics[width=84mm]
                    {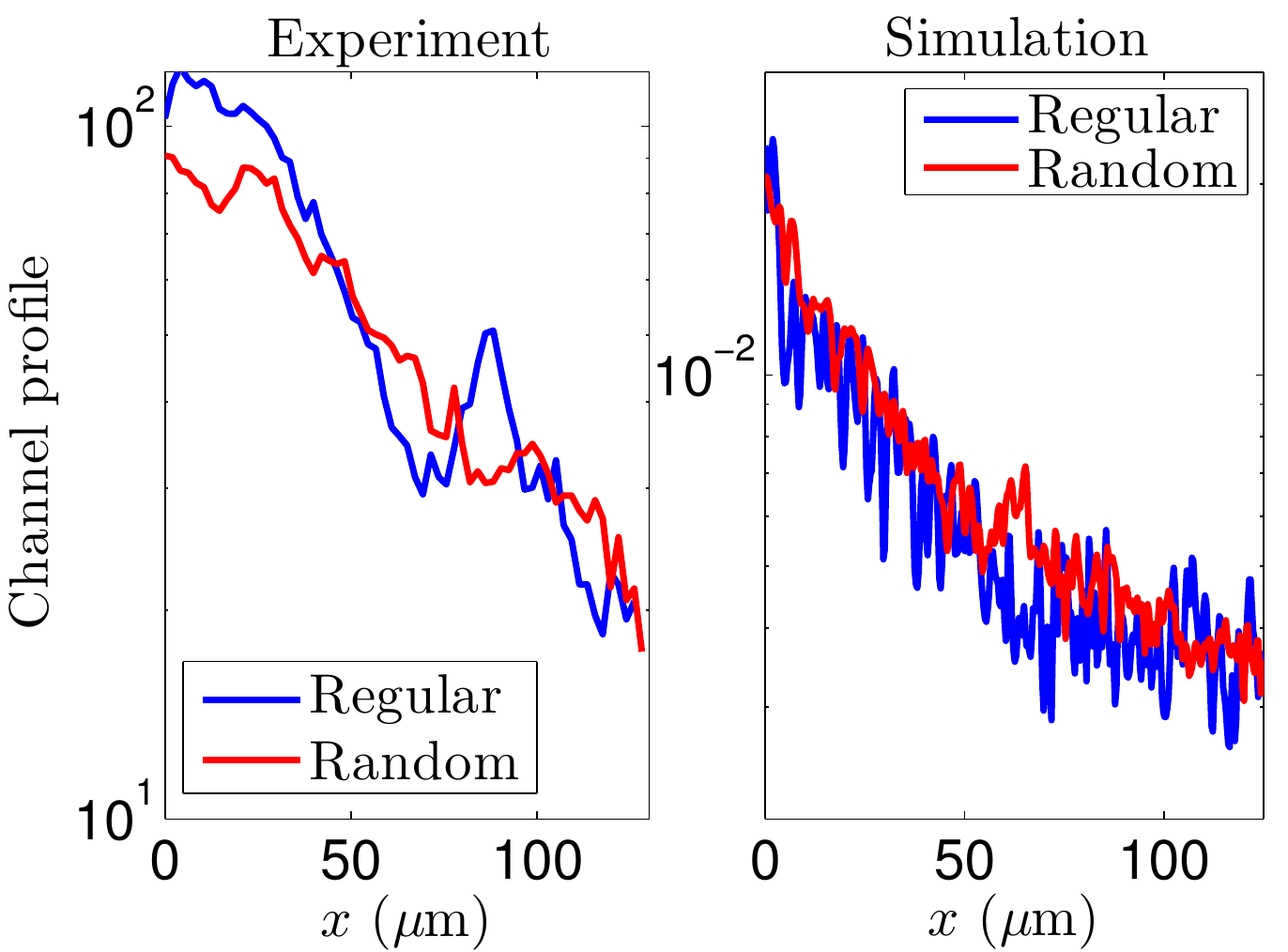}
    \caption{\label{fig:Supplementary_Ordered}
             \textbf{Long-time density profile in regular and random lattices}
             (a) Experimental data. The density at fill-factor $\eta=0.13$, $(r,L,w)=(43,162,43)~\mu$m is shown for scatterers
             arranged randomly, compared with scatterers arranged in a square
             regular lattice. (b) Numerical simulation, with equivalent parameters for (a).
            }
\end{figure}

For comparison, we use the tuneability of the spatial light modulator to compare a disordered system with an ordered system, in a regime of weak localisation. In
Fig.~\ref{fig:Supplementary_Ordered}(a), we compare the experimental density
profiles obtained when the same number of point scatterers ($\eta=0.13$) are
arranged randomly, compared to in a regular ordered square lattice. We observe
stronger linearity in the semilogarithmic plot for the disordered case. The $R^{2}$ value of the semilogarithmic profile in the regular channel is 0.90, compared with 0.95 of the disordered channel. We also point out the curvature observed in the first 20~$\mu$m in the case of regular scatterers. The increased exponential character of the density profile in the disordered lattice is also found in the simulation (Fig.~\ref{fig:Supplementary_Ordered}(b)), and the simulation indicates that the resistance of the disordered lattice ($1.7\times 10^{-3}h$) is 20\% larger than the regular lattice ($1.4\times 10^{-3}h$). We note the high-frequency spatial oscillation present in the simulated regular-lattice density profile, which we do not observe in the experiment due to the imaging resolution. A full study of the comparison between regularly arranged and random scatterers is currently in
preparation.

\section{Numerical simulation}

We provide the values of physical parameters used in the simulation in Table~\ref{tab:phys}, and the numerical parameters in Table~\ref{tab:num}. We solve the Gross-Pitaevskii equation
\begin{equation*}
    i \hbar\, \frac{\partial \psi(\mathbf{r}, t)}{\partial t}
    =
    \left \lbrack
        -\frac{\hbar^{2}}{2m} \nabla^2_{\text{2D}} +
        V_{\text{trap}}(\mathbf{r}, t)           +
        V_{\text{int}}(\mathbf{r}, t)
    \right \rbrack
    \psi(\mathbf{r}, t)
\end{equation*}
with initial condition $\psi(\mathbf{r}, t) = \psi_{0}(\mathbf{r})$. Here
$V_{\text{int}}(\mathbf{r})$ denotes the interaction potential
\begin{equation*}
    V_{\text{int}}(\mathbf{r})
    =
    g N \left\vert \psi(\mathbf{r}) \right\vert^{2}
    =
   \frac{2\sqrt{2\pi}\, \hbar^{2} a_{\mathrm{s}}}{ma_{\mathrm{z}}}\,
   N \left|\psi(\mathbf{r})\right|^{2},
\end{equation*}
with $a_{\mathrm{s}}$ being the $s$-wave scattering length of the \textsuperscript{87}Rb
atoms, $a_{\mathrm{z}}$ is the oscillator length of the harmonic oscillator corresponding
to $\omega_{\mathrm{z}}$. The trapping potential, $V_{\text{trap}}(\mathbf{r})$, includes
the strong dumbbell-shaped well of average depth $V_{\text{db}}$, the artificial
gravitational potential and a weak harmonic potential with minimum at the centre
of the dumbbell. We also note here that the presence of the linear tilt causing
the artificial gravitational potential, and of the weak harmonic trap in the
numerical simulation are included in the calculations for consistency with the
experimental setup. The interference fringes due to distortion of the 1064~nm
beam wavefront within the 2D trap are not modelled.

\begin{table}
    \begin{tabular}{lp{1mm}lp{1mm}l}
        Name                              & & Symbol           & & Value                            \\
        \hline\hline
        particle number                   & &  $N$             & & $ 16000$                         \\
        \textsuperscript{87}Rb mass       & &  $m$             & & $ 87 \times \text{amu}$          \\
        Trap frequency (x)                & &  $\omega_{\mathrm{x}}$    & & $ 2\pi \times 1 $ rad/s          \\
        Trap frequency (y)                & &  $\omega_{\mathrm{y}}$    & & $ 2\pi \times 1 $ rad/s          \\
        Trap frequency (z)                & &  $\omega_{\mathrm{z}}$    & & $ 2\pi \times 800$ rad/s         \\
        angular frequency of initial trap & &  $\omega_{0}$    & & $ 2\pi \times 25$  rad/s         \\
        $s$-wave scattering length        & &  $a_{\mathrm{s}}$         & & $ 107 a_{0}$                     \\
        nonlinearity length               & &  $\lambda$       & & $ 2 \sqrt{2\pi}\, N a_{\mathrm{s}}$       \\
        angle of effective gravity        & &  $\theta$        & & $ 0.0002^{\circ}$                \\
        potential depth                   & &  $V_{\text{db}}$ & & $ 22$\,nK                        \\
        scatterer strength                & &  $V_{\text{sc}}$ & & $ 5$\,nK
    \end{tabular}
    \caption{Physical parameters in the numerical simulation.}
\label{tab:phys}
\end{table}

\begin{table}[bh]
    \begin{tabular}{p{48mm}p{1mm}lp{1mm}l}
        Name                                                         & & Symbol  & & Value       \\
        \hline\hline
        Spatial extension of the numerical grid in the $x$ direction & & $L_{\mathrm{x}}$ & & $500\,\mu$m \\
        Spatial extension of the numerical grid in the $y$ direction & & $L_{\mathrm{y}}$ & & $225\,\mu$m \\
        Number of grid points \newline in the $x$ direction          & & $n_{\mathrm{x}}$ & & $1536$      \\
        Number of grid points \newline in the $y$ direction          & & $n_{\mathrm{y}}$ & & $768$
    \end{tabular}
    \caption{Parameters for the numerical simulation.}
\label{tab:num}
\end{table}
The initial wavefunction is the ground-state wavefunction of $N$ interacting
particles in a three-dimensional harmonic potential with angular frequencies
$\omega_{0}$. This ground-state wavefunction, $\psi_{0}(\mathbf{r})$, is
determined by the standard imaginary-time propagation method \cite{Bao2004,
Magnus1954}, and is shifted to the source reservoir. In most of the simulation
its centre is located at the opening of the channel. However, we note that the
exact position does not strongly influence the transport properties, and only
moderately affect the timing. Using the adaptive, fourth-order Runge-Kutta-Fehlberg
method\cite{Burden2011} we propagate this initial wavefunction (reduced to a 2D
wavefunction) in real time over a grid of size $n_{\mathrm{x}} \times n_{\mathrm{y}}$
representing the rectangular area $L_{\mathrm{x}} \times L_{\mathrm{y}}$ in real space. Here we mention that unlike the popular fourth order Runge--Kutta method (RK4),
the Runge--Kutta--Fehlberg method (RKF45) is an adaptive method, i.e., it chooses
the best step-size to meet a predefined error threshold. Therefore the time-step
varies during in our simulations to meet the error threshold of $7 \times
10^{-12}$ in $L_{2}$-norm of the wavefunction. The
dumbbell --for any channel length and circular reservoir radius-- is positioned
at the centre of this grid symmetrically, i.e, the centre of the channel is at
the centre of the numerical grid.

\section{Effect of interactions}

\begin{figure}[bth]
    \includegraphics[width=250px]{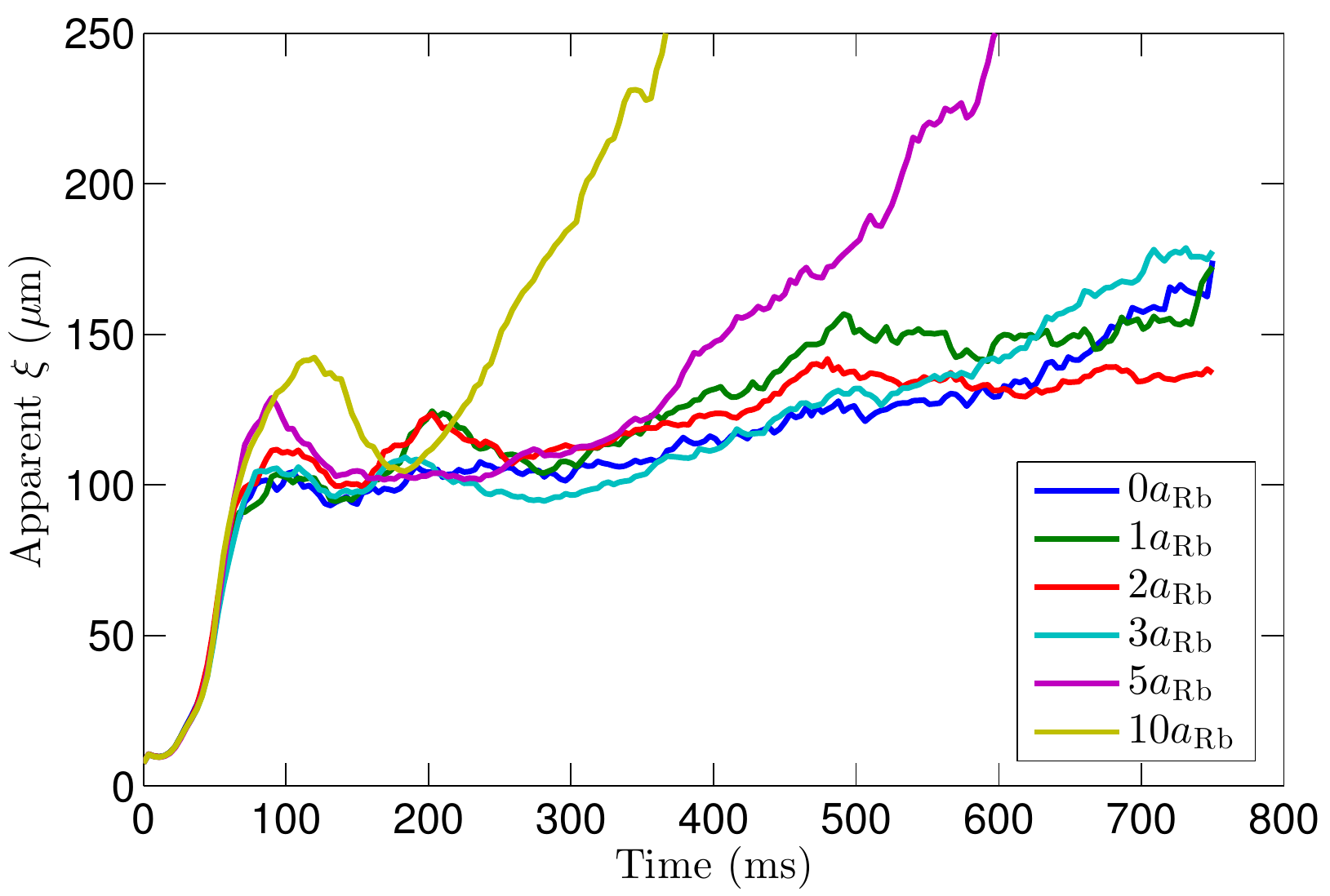}
    \caption{\label{fig:interaction}
             \textbf{Effect of interaction strength on localisation length.}
             Apparent localisation lengths as a function of time are plotted for
             a range of interaction strengths, with scattering lengths ranging
             from 0.0--10\,$a_{\mathrm{Rb}}$, as determined by numerical simulation. For this data, $(r,L,w)=(43,108,58)~\mu$m and $\eta=0.32$.
            }
\end{figure}

In Fig.~\ref{fig:interaction} we conduct numerical simulations for a range of
interaction strengths, to determine the effect of interactions on Anderson
localisation. The simulations are performed by allowing the atoms to expand from
the condensate with scattering length $a_{\mathrm{Rb}}$. Once the atoms have filled the
first reservoir, they have acquired their initial $k$-vector distribution, and
at this point in the simulation the interaction strength is abruptly changed to
a multiple of $a_{\mathrm{Rb}}$.

We observe no significant difference in the localisation lengths obtained with
scattering lengths 0 and $a_{\mathrm{Rb}}$, indicating that we may view the system as
near-non-interacting. This is due to the very low atomic density, averaging
approximately 1 atom / $\mu$m$^2$.

We observe that for $a\leq 4a_{\mathrm{Rb}}$, the measured apparent localisation length
tends to a steady state in time. For $a>6a_{\mathrm{Rb}}$, we observe that the channel
density profile does not tend to a steady localised state, and instead exhibits
a characteristic $\xi \sim \sqrt{t}$, indicative of diffusive
expansion. These results indicate that the experiment is conducted in a
weakly-interacting regime in which interactions do not significantly affect
Anderson localisation.

We note that the slow apparent increase in localisation length for $t>400$~ms is due in large part to atoms which re-enter the channel from the drain reservoir, and it does not imply a weakening of Anderson localisation.

\section{Relationship between \texorpdfstring{$\ls$}{ls} and \texorpdfstring{$\ltr$}{ltr}}

The elastic mean free-path between scattering events is approximately given by
the mean spacing between scatterers $\ls = \sigma/\sqrt{\eta}$, while the
transport mean free path, $\ltr$, also known as the Boltzmann mean free path, is
the distance over which the memory of the initial direction is lost. It is
found~\cite{Muller2005} that $\ltr = \Lambda(k\sigma)
\ls$, with a proportionality constant $\Lambda$ dependent on relative size of a
scatterer $\sigma$ and the de Broglie wavelength $\lambda = 2\pi/k$. Based on the data in Fig.~3(e) of the main text, $k\sigma\approx 1$ for the peak value of $k$ within the channel. We emphasise that
there is a distribution of atomic energies in our sample and that
$\Lambda$ depends strongly on $k \sigma$.

First let us determine a few characteristic quantities derived from classical or
semi-classical approximations. One length scale is provided by the average momentum of atoms within the channel, as found from Fig.~3(e) of the main text:
\begin{equation}
    \lambda_{\text{dB}}
    =
    \frac{2 \pi}{k}
    \approx
    3.9 \,\mu\text{m}.
\end{equation}
\noindent Within the channel, the de Broglie wavelength is significantly larger than the scatterer size $\sigma \approx
1.44\,\mu$m. The physical
system also possesses other characteristic length scales: the length of the
disordered channel $L_{0} \approx 36-180\,\mu$m, the channel width $w \approx
14-87\,\mu$m, and the mean minimal distance between scatterers $\ell_{\mathrm{s}} =
\sigma/\sqrt{\eta} \approx 2.5-3.4\,\mu$m for the corresponding fill-factors
$\eta = 0.32$ and 0.17, respectively. In general, therefore, their relationship in the sequence of our
experimental runs is $\sigma \lesssim \lsc < \lambda_{dB} \lesssim \ltr < w < \xi \lesssim L_{0}$.

As for length scales, there are also some characteristic energy scales of this
system which are given below as temperatures. Furthermore for all length scales
we may associate an energy scale as well via $\text{energy} \propto \hbar^{2}/m
(\text{length})^{2}$. We estimated the condensate temperature to be
$T_{\text{BEC}} \approx 5$\,nK, and the BEC is released in a dumbbell-shaped
potential with depth of $T_{\text{pot}} \approx 22$\,nK. The random scatterers
have a height of $T_{\text{sc}} \approx T_{\text{pot}} \approx 5$\,nK.
Out of these energy scales we note here the highest which corresponds to
$\sigma$, the shortest length scale: $E_{\sigma} = \hbar^{2}/m \sigma^{2}
\approx 2.7\,$nK.

The scattering process has a decisive parameter, $k \sigma$, i.e., the
relative size of the matter-wave compared to a single scatterer. Using the
approximation, $k \approx \kdB$, one obtains $\kdB \sigma \approx 2.2$. Such
value of $k \sigma$ suggests a non-isotropic scattering process even though we
are still in the weak scattering regime since the atoms kinetic energies are
higher than $E_{\text{sc}} = \eta E_{\sigma}$ for all $\eta$ values.

In order to establish a relationship between $\ltr$ and $\ls$ we evaluate
Eq.~(6) in Ref.~\cite{Muller2005}
\begin{equation}
    \label{eq:Supplementary_MeanFreePaths}
    \frac{1}{\Lambda(k\sigma)} =
    \frac{\lsc}{\ltr}
    =
    1 -
    \frac{\int_{0}^{2\pi}%
              {\cos{(\theta)} \mathcal{P}(k\sigma, \theta) d\theta}%
         }
         {\int_{0}^{2\pi}%
              {\mathcal{P}(k\sigma, \theta) d\theta}%
         }
\end{equation}
where $\mathcal{P}(k\sigma, \theta) = 8 \mathcal{F}\!\left ( k\sigma
\sin{\!( \theta/2 )}\right )$ and $\mathcal{F}(x) = \left
\lbrack \arccos{\!(x)}  - x \sqrt{1-x^{2}}\right \rbrack \Theta(1-x)$, while
$\Theta$ denotes the Heaviside distribution.

\begin{figure}
    \includegraphics[width=84mm]
                    {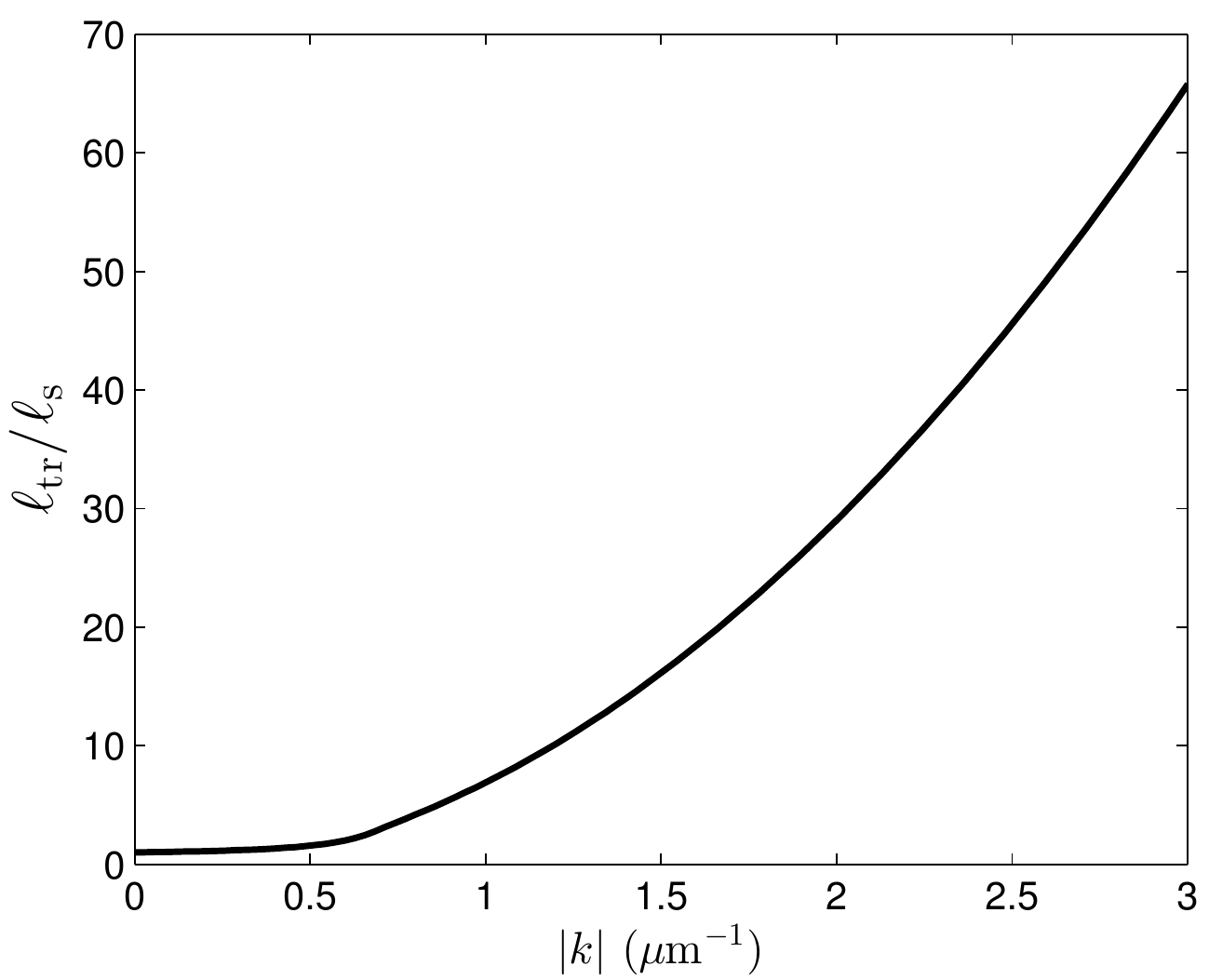}
    \caption{\label{fig:Supplementary_MeanFreePathsRatio}
             \textbf{$\boldsymbol{\Lambda}$ as a function of
                     wavenumber.}
             The ratio of the transport mean free path $\ell_{\mathrm{tr}}$ to the scattering mean free path $\ell_{\mathrm{s}}$ is plotted as a function of $|k|$, where $\sigma$ is assumed to be 1.4~$\mu$m. For $|k|=1.0~\mu$m$^{-1}$, $\ell_{\mathrm{tr}}\approx 7\ell_{\mathrm{s}}$. 
            }
\end{figure}

Figure~\ref{fig:Supplementary_MeanFreePathsRatio} shows the ratio of the transport mean free path $\ell_{\mathrm{tr}}$ to the scattering mean free path $\ell_{\mathrm{s}}$, as a function of the wavenumber $|k|$. As $k\sigma \rightarrow 0^{+}$ the
function $\mathcal{P}$ is more or less constant $4\pi$. In this limit one may
utilise that $\mathcal{F}(x) \sim \frac{\pi}{2} - 2x + \frac{1}{3} x^{3}$ and
determines the integrals in Eq.~\eqref{eq:Supplementary_MeanFreePaths}
analytically to obtain
\begin{equation*}
    \frac{\ltr}%
         {\lsc}
    \sim
    1 + \frac{8}{3\pi^{2}} (k\sigma) + \frac{256}{9\pi^{4}} (k\sigma)^{2}
    \qquad \text{as } (k\sigma) \rightarrow 0^{+}.
\end{equation*}
In the opposite limit, $k\sigma \rightarrow \infty$, the Heaviside distribution
in $\mathcal{F}$ is non-zero for $0 \le \theta \le 2 \arcsin{\!(1/k\sigma)}
\approx 2/k\sigma$ or $2\pi - 2/k\sigma \le \theta \le 2\pi$. Therefore the
bounds of integrals in Eq.~\eqref{eq:Supplementary_MeanFreePaths} are also
restricted to these two small intervals. However, the few leading terms in the
Taylor expansion of $\mathcal{F}$ are not sufficient to determine the asymptotic
behaviour of $\Lambda(k \sigma)$, but one needs to retain all terms in the
expansion
\begin{equation*}
    \mathcal{F}(x)
    =
    \frac{\pi}{2} - 2x +
    \sum_{n=1}^{\infty}%
        {\frac{1}{2^{2n-1} (2n-1)(2n+1)} \binom{2n}{n} x^{2n+1}}.
\end{equation*}
The leading term of $\Lambda(k \sigma)$ is quadratic, $c_{2} (k\sigma)^{2}$,
where the explicit expression of $c_{2}$ cannot be given in finite terms. Its
value is approximately
\begin{equation*}
    c_{2}
    \cong
    \frac{4}{3} \, \frac{6\pi-11}{6\pi-16}
    \approx
    3.673
\end{equation*}

At the end we may evaluate the analytic first order correction \cite{Muller2005}
to the Boltzmann diffusion coefficient for an indicative value of the
fill-factor, $\eta = 0.06$,
\begin{equation*}
    \frac{\delta D}{D_{\text{B}}}
    =
    \frac{2}{\pi}\,
    \frac{\ln{\!\left ( L_{0}/\ls \right )}}%
         {k \ltr}
    \approx 3.6 \times 10^{-3}.
\end{equation*}

%